\documentstyle[aps]{revtex}
\renewcommand{\Re}{{\mathrm{Re}}}
\renewcommand{\Im}{{\mathrm{Im}}}

\begin{document}
\title{Photo- and Electroproduction of Eta Mesons
\thanks{This work has been supported by the Deutsche
Forschungsgemeinschaft
(SFB 201).} \\
}
\author{G. Kn\"ochlein \and D. Drechsel \and L. Tiator \\
{\normalsize\it Institut f\"ur Kernphysik, Johannes
Gutenberg-Universit\"at, D-55099 Mainz, Germany}\\
{\small E-Mail: knoechlein@vkpmzp.kph.uni-mainz.de} }
Becherweg 45,
\maketitle
\begin{abstract}
Eta photo- and electroproduction off the nucleon is investigated
in an
effective lagrangian approach that contains Born terms and both
vector meson and nucleon resonance contributions. In particular,
we review and
develop the formalism for coincidence experiments with polarization
degrees
of freedom. The different response functions appearing in single and
double polarization experiments have been studied. We will present
calculations for structure functions and kinematical conditions
that are
most sensitive to details of the lagrangian, in particular with
regard to
contributions of nucleon resonances beyond the
dominant $S_{11}$(1535)
resonance.
\end{abstract}
\newpage

\section{Introduction}

The advent of new electron accelerators and intense photon
sources has
given rise to a renewed interest in photo-and electroproduction
of eta
mesons. Previous attempts of the late 60's to study this process
were
severely limited by large error bars
\cite{Del67,Bac68,ABB68,CBC68,Bl,He},
and also
a few later measurements
did not substantially improve the data
basis \cite{Dyt90,Hom88}.
The recent data of the
TAPS collaboration at MAMI have shown, however,
that this situation is
rapidly changing by use of modern accelerators and
detection equipment
\cite{Kr94}.
The high-quality data for angular distributions and total cross sections
for photon energies between threshold and 790 MeV may be considered  to be a
qualitative break-through in the experimental field. Similarly, data at the
higher energies up to 1150 MeV will be provided soon by the Phoenics
collaboration at ELSA \cite{Br94}.
At present, the data basis for electroproduction
of eta mesons is still very scarce. It is limited to a few older Bonn
data \cite{Ni78}
with large error bars and a more recent investigation of the Elan
collaboration at ELSA at very small momentum transfer, $Q^2 = 0.056
\, {\mathrm GeV}^2$
\cite{Wi93}.\\
Since eta production is strongly dominated by the resonance
$S_{11} (1535)$,
its coupling to other resonances is difficult to extract even from very
precise angular distributions. Instead, polarization observables will have
to play a major role in order to constrain the small resonance couplings.
Such experiments have been proposed, e.g., at the laser backscattering
facilities LEGS at Brookhaven and GRAAL at Grenoble. The experiments at
GRAAL will cover the energy range up to 2 GeV in the c.m. frame and provide a
first look at $\eta'$ production. Moreover, a series of experiments has
been planned at CEBAF to study both $\eta$ and $\eta'$ production by use
of polarization degrees of freedom \cite{CEBAF}.
\\
In contrast to the isovector pions which excite both $\Delta (T = 3/2)$ as
well as $N^{\ast} (T = 1/2)$ resonances, the isoscalar eta meson can
couple only to resonances with $T = 1/2$. However, a fascinating property
of the eta is its strong coupling to the $S_{11} (1535)$ and a very weak
coupling to all other $N^{\ast}$ resonances. Since the $S_{11} (1535)$ has
a 45 - 55\% branching ratio for eta decay, the eta is the ideal probe
to "tag" and investigate this resonance. In comparison the $S_{11} (1650)$
has
only a 1 \% branching ratio for eta decay. This fact has been
explained in  the quark model by a strong mixing between these two
resonances \cite{IK78}.
It is less obvious why none of the other resonances seems
to couple  strongly to the eta. In particular, the neighbouring dipole
excitation $D_{13} (1520)$ has a branching ratio of the order of $10^{-3}$.
The coupling of the Roper resonance $P_{11}(1440)$ to the eta meson is
difficult to determine, because the resonance mass is below threshold.
A previously reported strong coupling to the $P_{11} (1710)$ has been
dropped in the last issue of the Particle Data Group \cite{PDG94},
because the presented
evidence is not convincing. Further resonances with branching ratios of the
order of 1
\% or less are the states $D_{15}(1675)$, $D_{13}(1700)$ and $P_{13}(1720)$
\cite{PDG94,PDG92}.
All of these states
appear as resonances in $\pi N$ phase shift analyses, but precisely the
$S_{11}(1535)$ is not a good candidate for a resonance in the $\pi N$
system. The speed-plot of the $\pi N$ $S$-wave phase shift shows a very
untypical
form for a resonance \cite{hoe1,hoe2},
probably because of a strong cusp effect due to
the coupling to the eta channel.\\
$SU(3)$ flavour symmetry relates the coupling constant between the eta and
the nucleon to the pion-nucleon interaction,
\begin{equation}
\frac{g^2_{\eta NN}}{4\pi} = \frac{1}{3} (3 - 4 \frac{D}{D+F})^2\frac{g^2_
{\pi NN}}
{4\pi} \; .
\end{equation}
The resulting value ranges between 0.8 and 1.9 and depends strongly on
the strength parameters  F and D of the two
types of octet meson-baryon couplings. A previous analysis has shown,
however, that much smaller values of the order of 0.4 are necessary to fit
the data for eta photoproduction \cite{TBK94}. In the same analysis, the
forward-backward asymmetry of the angular distribution \cite{Kr94}
gave rise to the
interesting observation that pseudoscalar (PS) coupling is to be favoured
over pseudovector (PV) coupling. This finding seems to contrast the
derivation of low energy theorems (LET), which require PV coupling for
reasons of chiral invariance. However, such theorems are probably
irrelevant for eta production because of the following reasons. LET
describes the dominant $S$-wave multipole $E_{0+}$ by a power series in the
ratio of eta and nucleon mass, $m_{\eta}/m_N$. Obviously, this expansion is
not expected to converge well. Even in the much more favourable case of
pion photoproduction with $\mu = m_{\pi}/m_N\approx 1/7$, the low-energy
limit of $E_{0+}$ is difficult to predict. As has
been shown in detailed calculations within the framework of chiral
perturbation theory \cite{Mei},
higher loop effects play a major role and even a
calculation up to terms in $\mu^3$ does not seem to be sufficient. Second,
the $S_{11}(1535)$ is close to threshold and dominates the cross section.
In view of such a nearby singularity in the complex energy plane, a power
series expansion in $m_\eta$ is not expected to work near eta threshold.
Third, the physical $\eta$ and $\eta'$ mesons are combinations of the
corresponding particles in the $SU(3)$ octet and singlet with a mixing angle
of about $-20^{\circ}$. In this way the $U_A$(1) anomaly of the axial
singlet current will also affect the eta meson, and the corresponding
axial current will not be conserved, even in the Goldstone limit of
massless particles. \\
As has been pointed out, the slight backward peaking of the angular
distributions can be explained by PS coupling between eta and nucleon,
while a calculation with PV coupling would lead to a forward peaked cross
section. The difference between the two schemes is due to a sign change in
the $P$-wave $M_{1-}$ multipole. This multipole would also be affected by a
sizeable eta branch of the $N^{\ast}(1440)$ or Roper resonance. Though
there are no indications of such an effect at present, a careful study of
the smaller multipoles in eta photoproduction is certainly well
motivated. \\
Previous attempts to describe eta photoproduction have involved Breit-Wigner
functions for the resonances \cite{Hicks} or relativistic lagrangians to
describe resonant and nonresonant contributions \cite{BM91}. In a basically
nonrelativistic approximation, coupled-channel calculations have been
developed to treat the reactions $\pi N\rightarrow\pi N$,
$\pi N\rightarrow\pi\pi N$ and $\pi N\rightarrow\eta N$ in a unitary way
\cite{BT91,Sauer,Feu}.
Unfortunately the corresponding 3$\times$3 $S$-matrix of the hadronic
sector is not sufficiently well determined by the existing data, which
leaves a considerable model dependence. It is interesting to see that a
generalized Lie model can describe the hadronic data perfectly well, both
with and without an explicit resonance \cite{Den94}. In the latter case
the resonance-like structure of the data is built up by a cusp effect
due to the strong
$\pi-\eta$ coupling. Though the two models lead to completely different
singularities in the complex energy plane, the present data are not
sufficient to discard one or the other option.\\
In the following calculations we will follow the model of Bennhold and
Tanabe \cite{BT91},
which has already been applied to eta photoproduction in the
region of the $S_{11}(1535)$. We will extend the model to higher energies
and include higher resonances in a phenomenological way. Moreover, we will
study the potential of polarization degrees of freedom in both
photo- and electroproduction. It is  our main aim to select
kinematical situations
and polarization observables most favourable for identifying higher
multipole contributions in the presence of the strong $S$-wave transition.\\
In Sect. 2 we will outline the formalism to describe photo- and
electroproduction up to the most general polarization experiment.
The ingredients of a phenomenological model for eta production will be
discussed in Sect. 3. Our results for photo- and electroproduction
observables will be presented in Sects. 4 and 5, respectively, followed
by a brief conclusion in Sect. 6.

\section{Formalism and Polarization Degrees of Freedom}

To lowest order of the fine structure constant
$\alpha = e^2/4 \pi \approx 1/137$ the electron interacts with the currents
of the hadronic system through the exchange of one virtual photon
with well defined energy, $\omega = \varepsilon_f - \varepsilon_i$,
and momentum transfer, $\vec q = \vec k_f - \vec k_i$.
The Lorentz vectors $k_i = ( \varepsilon_i, \vec k_i)$ and
$k_f = ( \varepsilon_f, \vec k_f)$ characterize the kinematics of
the incident
and outgoing electron. In contrast to a photoproduction
experiment, where photon energy and momentum are related by
$ q^2 = \omega^2 - \vec q^2 = 0 $, the virtual photon offers the
possibility of an independent variation
 of photon energy and momentum, leading to information on the
spatial structure of the hadronic system.
As a consequence the kinematical situation in an electron scattering
experiment (see also Fig. \ref{kinpic}) is determined by three
independent kinematical variables, which
are usually chosen to be the invariant energy $W$,
the scattering angle
in the c.m. system $\Theta$ and the momentum
transfer $Q^2 = - q^2$. In a photoproduction
experiment only two kinematical variables are required to
describe the kinematics uniquely. In this case the
usual choice of variables is the scattering angle $\Theta$ and the photon
energy in the laboratory system $\nu$.

The nucleon is characterized by the Lorentz vector
$P_i = (E_i, \vec P_i)$
and
$P_f = (E_f, \vec P_f)$
in the initial and final state, respectively, and the 4-momentum of the
eta meson is denoted by
$k = ( \omega_{\eta}, \vec k)$.
Of course, the whole formalism presented here can be easily
applied to other pseudoscalar
meson electroproduction experiments, e.g. pion
production.
The threshold photon energy
for eta production is
\begin{equation}
\nu_{thr} = \frac{2 m_N m_{\eta} + m_{\eta}^2 + Q^2}{2 m_N} \; .
\end{equation}
Using the notation \cite{BD64}, the differential cross section for
the electroproduction process can be written as
\begin{eqnarray}
d \sigma & = & \frac{\varepsilon_i}{k_i} \frac{m_e}{\varepsilon_i}
\frac{m_N}{E_i} \frac{m_e}{\varepsilon_f} \frac{d^3 k_f}{(2 \pi)^3}
\frac{1}{2 \omega_{\eta}} \frac{d^3 k}{(2 \pi)^3} \frac{m_N}{E_f}
\frac{d^3 P_f}{(2 \pi)^3}
\nonumber \\
 & & \times (2 \pi)^4
\delta^{(4)} ( P_i + q - k - P_f )
\nonumber \\
& & \times \mid \langle P_f, k \mid J^{\mu} \mid P_i \rangle q^{-2}
\langle k_f \mid j_{\mu} \mid k_i \rangle \mid^2 \; ,
\end{eqnarray}
where the phase space is evaluated in the laboratory frame, and $j^{\mu}$
and $J^{\mu}$ denote the electromagnetic currents of the electron and the
hadronic system, respectively.

The most general lorentz- and
 gauge invariant current operator $J^{\mu}$ for the production of a
pseudoscalar particle off a nucleon between nucleon Dirac spinors is given
by the CGLN parametrization \cite{CGLN} of the transition current,
\begin{eqnarray}
\label{CGLNcurr}
\vec J & = & \frac{4 \pi W}{m} \left[ i
\mbox{\boldmath{$\tilde{\sigma}$}} F_1
            + \left( \mbox{\boldmath{$\sigma$}} \cdot
              {\hat{\vec k}} \right)
              \left(
\mbox{\boldmath{$\sigma$}} \times {\hat{\vec q}} \right) F_2
            + i \tilde{ \vec k} \left(
\mbox{\boldmath{$\sigma$}} \cdot \hat{ \vec q} \right) F_3 \right.
\nonumber \\
        & & \left. + i \tilde{\vec k} \left(
\mbox{\boldmath{$\sigma$}} \cdot \hat{ \vec k} \right) F_4
            + i \hat{\vec q} \left( \mbox{\boldmath{$\sigma$}} \cdot
\hat{\vec q}
\right) F_5
            + i \hat{\vec q} \left( \mbox{\boldmath{$\sigma$}} \cdot
\hat{\vec k} \right) F_6  \right] \; , \\
\label{CGLN2}
\rho & = & \frac{4 \pi W}{m} \left[ i \left(
\mbox{\boldmath{$\sigma$}} \cdot \hat{\vec k} \right) F_7
           + i \left( \mbox{\boldmath{$\sigma$}} \cdot \hat{\vec q} \right)
F_8 \right] = \frac{\vec q \cdot \vec J}{\omega} \; ,
\end{eqnarray}
with all vectors expressed in the c.m. frame and
$\tilde{\vec \sigma} = \vec \sigma - ( \vec \sigma \cdot \hat{\vec q})
\hat{\vec q}$ etc. In this parametrization
the first four terms represent the purely
transverse component, $\vec J_{T}$,
of the current $\vec J$, while the fifth and sixth term
 are the longitudinal part, $\vec J_L$.
The longitudinal component of the current is related to the scalar
density $\rho$ via current conservation, $q_{\mu} J^{\mu}=0$,
and, as a consequence,
\begin{equation}
 \mid \vec q \mid F_5 = \omega F_8 \; , \, \, \,
\mid \vec q \mid F_6 = \omega F_7 \; .
\end{equation}
The CGLN amplitudes
are complex functions of three independent
kinematical variables, $F_i = F_i ( W, \Theta, Q^2 )$.
The degree of transverse
polarization of the virtual photon,
\begin{equation}
\varepsilon = \left( 1 + \frac{2 {\vec q}^{\,2}}{Q^2} \tan^2
\frac{\Theta_e}{2}
\right)^{-1}\; ,
\end{equation}
is invariant under collinear transformations, and
$\vec q$ and $\Theta_e$ may be expressed in the lab or c.m. frame,
while the
longitudinal polarization,
\begin{equation}
\varepsilon_L = \frac{Q^2}{\omega^2} \varepsilon \; ,
\end{equation}
is frame-dependent.
Using standard techniques (see e.g. \cite{DR} or \cite{DT92}),
the differential cross
section for an electroproduction experiment is
\begin{eqnarray}
\frac{d \sigma}{d \Omega_f d \varepsilon_f d \Omega_{\eta}} & = & \Gamma
\frac{d \sigma_v}{d \Omega_{\eta}} \; ,
\end{eqnarray}
where the flux of the virtual photon field is given by
\begin{equation}
\Gamma = \frac{\alpha}{2 \pi^2} \frac{\varepsilon_f}{\varepsilon_i}
\frac{k_{\gamma}^{lab}}{Q^2} \frac{1}{1 - \varepsilon} \; ,
\end{equation}
and the virtual differential cross section is
\begin{eqnarray}
\frac{d \sigma_v}{d \Omega_{\eta}} & = &
\frac{d \sigma_T}{d \Omega_{\eta}}
+ \varepsilon_L \frac{d \sigma_L}{d \Omega_{\eta}}
+ \left[ 2 \varepsilon_L \left( 1 + \varepsilon \right) \right]^{1/2}
  \frac{d \sigma_{TL}}{d \Omega_{\eta}} \cos \phi_{\eta}
\nonumber \\
& & + \varepsilon \frac{d \sigma_{TT}}{d \Omega_{\eta}} \cos 2 \phi_{\eta}
+ h \left[ 2 \varepsilon_L \left( 1 - \varepsilon \right)
\right]^{1/2} \frac{d \sigma_{TL'}}{d \Omega_{\eta}} \sin \phi_{\eta}
\nonumber \\
& & + h \left( 1 - \varepsilon^2 \right)^{1/2} \frac{d \sigma_{TT'}}
{d \Omega_{\eta}} \; .
\end{eqnarray}
In this expression $k_{\gamma}^{lab} = \left( W^2 - m_i^2 \right) / 2 m_i $
denotes the ''photon equivalent energy'',
the laboratory energy necessary for a real photon to excite a hadronic
system with c.m. energy $W$.
Note that all kinematical variables appearing in the virtual photon cross
section $d \sigma_v / d \Omega_{\eta}$ have to be expressed in the c.m.
frame.
The individual contributions $d \sigma_{i}
/ {d \Omega_{\eta}}$ ($i = T, L, TL, TT, TL', TT'$)
are usually parametrized in terms of response functions which depend on the
independent kinematical variables.

Experiments with three types of
polarization can be performed in eta production:
electron beam polarization,
polarization of the target nucleon and polarization of
the recoil nucleon.
Target polarization will be described in the frame $\{ x, y, z \}$ in Fig.
\ref{polkoordinaten},
with the $z$-axis pointing
into the direction of the photon momentum $\hat{ \vec q}$, the
$y$-axis perpendicular to the reaction plane,
${\hat{ \vec y}} = {\hat{ \vec q}} \times {\hat{ \vec k}} /
\sin \Theta_{\eta}$,
and the $x$-axis given by ${\hat{\vec x}} = {\hat{ \vec y}}
\times {\hat{\vec z}}$.
For recoil polarization we will use
the frame $\{ x', y', z' \}$, with
the $z'$-axis defined by the momentum vector of the outgoing
eta meson ${\hat{\vec k}}$, the $y'$-axis as for target
polarization and the $x'$-axis given by
${\hat{\vec{x}'}} = {\hat{\vec{y}'}} \times {\hat{\vec{z}'}}$.
The most general expression for a coincidence experiment considering all
three types of polarization is
\begin{eqnarray}
\label{dsigmafull}
\frac{d\sigma_v}{d\Omega_{\eta}} & = &
\frac{\mid \vec k \mid}{k_{\gamma}^{cm}}
P_{\alpha} P_{\beta}
\{
R_T^{\beta \alpha}
+ \varepsilon_L R_L^{\beta \alpha} \nonumber \\
& & + \left[ 2 \varepsilon_L \left( 1 + \varepsilon \right)
\right]^{1/2}
( ^c \! R_{TL}^{\beta \alpha}
\cos \phi_{\eta}
+ ^s \! \! R_{TL}^{\beta \alpha}
\sin \phi_{\eta})
 \nonumber \\
& &
+ \varepsilon (
^c \! R_{TT}^{\beta \alpha}
\cos 2 \phi_{\eta}
+ ^s \! \! R_{TT}^{\beta \alpha}
\sin 2 \phi_{\eta}) \nonumber \\
& &
+ h \left[ 2 \varepsilon_L ( 1 - \varepsilon ) \right]^{1/2}
( ^c \! R_{TL'}^{\beta \alpha}
\cos \phi_{\eta}
+ ^s \! \! R_{TL'}^{\beta \alpha}
\sin \phi_{\eta}) \nonumber \\
& &
+ h ( 1 - \varepsilon^2 )^{1/2} R_{TT'}^{\beta \alpha}
\},
\end{eqnarray}
where $P_{\alpha} = (1, \vec P)$
and
$P_{\beta} = (1, \vec P')$.
Here $\vec P = (P_x, P_y, P_z)$ denotes the target
and $\vec P' = (P_{x'}, P_{y'}, P_{z'})$ the
recoil polarization vector.
The zero components $P_0 = 1$ lead to contributions in the cross section
which are present in the polarized as well as the unpolarized case.
In an experiment without target and recoil
polarization $\alpha = \beta = 0$, so
the only remaining contributions are $R_i^{00}$.
The functions $R_i^{\beta \alpha}$ describe the response of the hadronic
system in the process. Summation over Greek indices is implied.
An
additional superscript $s$ or $c$ on the left indicates a sine or cosine
dependence of the respective contribution on the azimuthal angle. Some
response functions vanish identically (see Table \ref{elprodtab}
for a systematic
overview ). The number of different response functions is further reduced
by the equalities listed in App. \ref{identresp}, and in the most general
electroproduction experiment 36 polarization observables can be determined.
The response functions $R^{\beta \alpha}_i$ are
real or imaginary parts of bilinear forms of the CGLN amplitudes
depending on the scattering angle $\Theta$.

\begin{table*}[ht]
\caption[Polarization Observables in Meson Electroproduction]
{
\label{elprodtab}
\small{
Polarization observables in pseudoscalar meson electroproduction.
{\small A star denotes a response function which does not vanish but is
identical to another response function
via a relation in App. \ref{identresp}.}
}
}
\renewcommand{\arraystretch}{1.3}
\begin{tabular}{c c ccc ccc ccccccccc}
\hline\noalign{\smallskip}
& & \multicolumn{3}{c}{Target} & \multicolumn{3}{c}{Recoil}
& \multicolumn{9}{c}{Target + Recoil} \\
\hline
$\beta$ & $-$ & $-$ & $-$ & $-$ & $x'$ & $y'$ & $z'$ & $x'$
& $x'$ & $x'$ & $y'$ &
$y'$ & $y'$ & $z'$ & $z'$ & $z'$ \\
$\alpha$ & $-$ & $x$ & $y$ & $z$ & $-$ & $-$ & $-$ & $x$
& $y$ & $z$ & $x$ & $y$ & $z$
& $x$ & $y$ & $z$ \\
\hline
$T$ & $R_T^{00} $ & $0$ & $R_T^{0y} $ & $0$ & $0$ & $R_T^{y'0}$
& $0$ &
$R_{T}^{x' x}$ & $0$ & $R_{T}^{x' z}$ & $0$ & $*$ & $0$ & $R_{T}^{z' x}$ &
$0$ & $R_{T}^{z' z}$ \\
$L$ & $R_L$ & $0$ & $R_L^{0y}$ & $0$ & $0$ & $*$ & $0$ &
$R_{L}^{x' x}$ & $0$ & $R_{L}^{x' z}$ & $0$ & $*$ & $0$ & $*$ & $0$
& $*$ \\
${ }^c TL$ & ${}^c R_{TL}^{00}$ & $0$ & $ { }^c R_{TL}^{0y}$
& $0$ & $0$ & $*$
& $0$ & $ { }^c R_{TL}^{x'x}$ & $0$ & $*$  &
$0$ & $*$ & $0$ & $ { }^c R_{TL}^{z'x} $ & $0$ & $*$ \\
${ }^s TL $ & $0$ & $ { }^s R_{TL}^{0x} $ & $0$ & $ { }^s R_{TL}^{0z} $ &
$ { }^s R_{TL}^{x'0} $ & $0$ & ${ }^s R_{TL}^{z'0} $ & $0$ & $*$ &$0$ &
$*$ &$0$ & $*$ &$0$ & $*$ & $0$ \\
${ }^c TT$ & ${}^c R_{TT}^{00}$ & $0$ & $*$ & $0$ & $0$ & $*$
& $0$ & $*$ & $0$ & $*$ &
$0$ & $*$ & $0$ & $*$ & $0$ & $*$ \\
${ }^s TT $ & $0$ & $ { }^s R_{TT}^{0x} $ & $0$ & $ { }^s R_{TT}^{0z} $ &
$ { }^s R_{TT}^{x'0} $ & $0$ & ${ }^s R_{TT}^{z'0} $ & $0$ & $*$ &$0$ &
$*$ &$0$ & $*$ &$0$ & $*$ & $0$ \\
${ }^c TL' $ & $0$ & $ { }^c R_{TL'}^{0x} $ & $0$ & $ { }^c
R_{TL'}^{0z} $ &
$ { }^c R_{TL'}^{x'0} $ & $0$ & ${ }^c R_{TL'}^{z'0} $ & $0$ & $*$ &$0$ &
$*$ &$0$ & $*$ &$0$ & $*$ & $0$ \\
${ }^s TL'$ & ${}^s R_{TL'}^{00}$ & $0$ & ${ }^s R_{TL'}^{0y}$
& $0$ & $0$ & $*$
& $0$ & ${ }^s R_{TL'}^{x'x}$ & $0$ & $*$ &
$0$ & $*$ & $0$ & ${ }^s R_{TL'}^{z'x}$ & $0$ & $*$ \\
$TT' $ & $0$ & $ R_{TT'}^{0x} $ & $0$ & $ R_{TT'}^{0z} $ &
$ R_{TT'}^{x'0} $ & $0$ & $R_{TT'}^{z'0} $ & $0$ & $*$ &$0$ &
$*$ &$0$ & $*$ &$0$ & $*$ & $0$ \\
\noalign{\smallskip}
\hline
\end{tabular}
\end{table*}

In photoproduction the longitudinal components vanish, and the
relevant response functions will be divided by the
transverse response function $R_T^{00}$ in order to obtain
the polarization observables. The common
descriptors of these observables
(see also \cite{Barker} and \cite{FTS92})
can be found in Table \ref{phprodtab}, the relation to the response
functions in App. \ref{photoelektro}.

\begin{table*}[ht]
\caption[Polarization Observables in Meson Photoproduction]
{
\label{phprodtab}
\small{Polarization observables in meson photoproduction. The entries in
brackets denote polarization observables, which also appear elsewhere in
the table.}
}
\renewcommand{\arraystretch}{1.3}
\begin{tabular}{c c ccc ccc cccc}
\hline\noalign{\smallskip}
 Photon & & \multicolumn{3}{c}{Target} & \multicolumn{3}{c}{Recoil}
& \multicolumn{4}{c}{Target + Recoil} \\
\hline
 & $-$ & $-$ & $-$ & $-$ & $x'$ & $y'$ & $z'$ & $x'$ & $x'$ & $z'$
& $z'$  \\
 & $-$ & $x$ & $y$ & $z$ & $-$ & $-$ & $-$ & $x$ & $z$ & $x$ & $z$ \\
\hline
transverse unpolarized ($T$) & $\sigma_0$ & $0$
& $T$ & $0$ & $0$ & $P$ & $0$ &
$T_{x'}$ & $-L_{x'}$ & $T_{z'}$ & $L_{z'}$ \\
transverse (linear pol.) ($TT$) & $- \Sigma$ & $H$ & $(-P)$ & $-G$ &
$O_{x'}$ & $(-T)$ & $O_{z'}$ &
$(-L_{z'})$ & $(T_{z'})$ & $(-L_{x'})$ & $(-T_{x'})$ \\
transverse (circular pol.) ($TT'$) & $0$ & $F$ & $0$ & $-E$ &
 $-C_{x'}$ & $0$ & $-C_{z'}$ &
$0$ & $0$ & $0$ & $0$ \\
\noalign{\smallskip}
\hline
\end{tabular}
\end{table*}

In contrast to electroproduction, there are no new
polarization observables in photoproduction accessible by
triple polarization experiments
(beam + target + recoil polarization).
As a consequence we may classify the differential cross sections
by the three classes of double
polarization experiments:
\begin{itemize}
\item polarized photons and polarized target
\end{itemize}
\begin{eqnarray}
\frac{d \sigma}{d \Omega_{\eta}} & = & \sigma_0
\left\{ 1 - P_T \Sigma \cos 2 \varphi \right. \nonumber \\
& & + P_x \left( - P_T H \sin 2 \varphi + P_{\odot} F \right)
\nonumber \\
& & - P_y \left( - T + P_T P \cos 2 \varphi \right) \nonumber \\
& & \left. - P_z \left( - P_T G \sin 2 \varphi + P_{\odot} E \right)
\right\}
\; ,
\end{eqnarray}
\begin{itemize}
\item polarized photons and recoil polarization
\end{itemize}
\begin{eqnarray}
\frac{d \sigma}{d \Omega_{\eta}} & = & \sigma_0
\left\{ 1 - P_T \Sigma \cos 2 \varphi \right. \nonumber \\
& & + P_{x'} \left( - P_T O_{x'} \sin 2 \varphi - P_{\odot} C_{x'} \right)
\nonumber \\
& & - P_{y'} \left( - P + P_T T \cos 2 \varphi \right) \nonumber \\
& & \left. - P_{z'} \left( P_T O_{z'} \sin 2 \varphi
+ P_{\odot} C_{z'} \right) \right\} \; ,
\end{eqnarray}
\begin{itemize}
\item polarized target and recoil polarization
\end{itemize}
\begin{eqnarray}
\frac{d \sigma}{d \Omega_{\eta}} & = & \sigma_0
\left\{ 1 + P_{y'} P
+ P_x \left( P_{x'} T_{x'} + P_{z'} T_{z'} \right) \right. \nonumber \\
& & \left. + P_{y} \left( T + P_{y'} \Sigma \right)
- P_{z} \left( P_{x'} L_{x'} - P_{z'} L_{z'} \right) \right\} \; .
\end{eqnarray}
In these equations $\sigma_0$ denotes the unpolarized differential cross
section, the
transverse degree of photon polarization is denoted by $P_T$,
$P_{\odot}$ is the right-handed circular photon polarization and
$\varphi$ the angle between photon polarization vector and
reaction plane.

In conclusion there are 16 different
polarization observables for real photon experiments. In
electroproduction there are four additional
observables for the exchange of longitudinal photons and sixteen
observables due to longitudinal-transverse interference. However,
as has become clear from the general form of the
transition current $J^{\mu}$,
there are only six independent complex
amplitudes uniquely describing the electroproduction process.
This
corresponds to six absolute values and five relative phases between the
CGLN amplitudes,
i.e. there are only eleven independent quantities
which completely and uniquely
determine the transition current $J^{\mu}$.

The aim of a so-called
''complete experiment'' is the
determination of the current $J^{\mu}$ for the
process under investigation in a given kinematics.
In order to carry through
such a complete experiment, it should
suffice to determine eleven independent
response functions.
The ambitious program of a complete experiment was formulated by
Barker et al. in the sixties for the photoproduction case \cite{Barker}.
Although it would be interesting to pursue such a project
for eta production, we will concentrate in the following on the
more physical aspects of this process.
In particular our theoretical investigations will supply information on
an adequate selection of response functions and observables. In the case of
eta production it will be of special interest to study response
functions which give clear information of the eta meson coupling to higher
resonances.

Another useful parametrization of the hadronic
current $\vec J$ is based on a decomposition into spherical
components $J_{\pm, 0}$, with
$ J_{\pm} = \mp (J_x \pm i J_y) / \sqrt{2} $ and $J_0 = J_z$. Since the
components of the current operator $\vec J$ are operators in
two-dimensional Pauli spinor space, we can represent them by
$2 \times 2$ matrices in terms of the
helicity amplitudes $H_i$:
\begin{eqnarray}
\label{heldef}
J_+ & = &
\left(
\begin{array}{cc}
H_1 & H_2 \\
H_3 & H_4
\end{array}
\right) \; , \; \;
J_- =
\left(
\begin{array}{cc}
H_4 & -H_3 \\
-H_2 & H_1
\end{array}
\right) \; , \nonumber \\
J_0 & = &
\left(
\begin{array}{cc}
H_5 & H_6 \\
H_6 & -H_5
\end{array}
\right) \; .
\end{eqnarray}
The response functions are expressed
by these helicity
amplitudes in App. \ref{helapp}. It is straightforward to calculate
the relation between CGLN and helicity amplitudes by comparing the
matrix elements of the current operator (\ref{CGLNcurr})
with the definitions
(\ref{heldef})
transformed into the Cartesian basis. The resulting transformations
are
\begin{eqnarray}
H_1 & = & \frac{-1}{\sqrt{2}} \sin \Theta (F_3 + F_4 \cos \Theta), \\
H_2 & = & \frac{-1}{\sqrt{2}} (2 F_1 - 2 F_2 \cos \Theta + F_4 \sin^2
\Theta), \\
H_3 & = & \frac{-1}{\sqrt{2}} F_4 \sin^2 \Theta, \\
H_4 & = & \frac{1}{\sqrt{2}} \sin \Theta (2 F_2 + F_3 + F_4 \cos \Theta), \\
H_5 & = & F_5 + F_6 \cos \Theta, \\
H_6 & = & F_6 \sin \Theta.
\end{eqnarray}

Electroproduction data are usually analyzed in terms of multipoles
$E_{l \pm}$, $M_{l \pm}$ and $L_{l \pm}$
characterizing the excitation mechanism
(electric ($E$), magnetic ($M$) or longitudinal ($L$))
and the orbital ($l$) and
 total angular momentum, which is given by $j = l \pm \frac{1}{2}$.
The relation between these meson production multipoles and
the type of electromagnetic multipole radiation $EL$, $ML$, $CL$
responsible for
resonance excitation can be read off from Table \ref{respartab}.
The connection with the CGLN amplitudes can be established via a multipole
series in terms of derivatives of the Legendre polynomials $P_l$
\cite{Am70},
\begin{eqnarray}
F_1 & = & \sum_{l \ge 0} \left\{ \left( M_{l+} + E_{l+} \right) {P'}_{l+1}
\nonumber \right. \\
& & \left. + \left[ \left( l+1 \right) M_{l-} +
E_{l-} \right] {P'}_{l-1} \right\} \;, \\
F_2 & = & \sum_{l \ge 1} \left[ \left( l + 1 \right) M_{l+}
+ l M_{l-} \right] {P'}_{l} \; , \\
F_3 & = & \sum_{l \ge 1} \left[ \left( E_{l+} - M_{l+} \right) {P''}_{l+1}
\right. \nonumber \\
& & \left. + \left( E_{l-} + M_{l-} \right) {P''}_{l-1} \right] \; , \\
F_4 & = & \sum_{l \ge 2} \left( M_{l+} - E_{l+} - M_{l-} - E_{l-} \right)
{P''}_{l} \; , \\
F_5 & = & \sum_{l \ge 0} \left[ \left( l + 1 \right) L_{l+} {P'}_{l+1}
- l L_{l-} {P'}_{l-1} \right] \; , \\
F_6 & = & \sum_{l \ge 1} \left[ l L_{l-} - (l + 1) L_{l+} \right] {P'}_{l}
\; .
\end{eqnarray}
The inversion of this set of equations is very useful for constructing
phenomenological models:
\begin{eqnarray}
E_{l+} & = & \int_{-1}^{1} dx
\left[ \frac{1}{2 (l + 1)} P_l F_1 - \frac{1}{2 ( l + 1 )} P_{l+1} F_2
\right.
\nonumber \\
& & \left. + \frac{1}{2 (l+1)} \frac{l}{2 l + 1} \left( P_{l-1} - P_{l+1}
\right) F_3 \right. \nonumber \\
& & + \left. \frac{1}{2 ( 2 l + 3 )} (P_l - P_{l+2} ) F_4 \right] \; , \\
E_{l-} & = & \int_{-1}^{1} dx
\left[ \frac{1}{2 l} P_l F_1 - \frac{1}{2 l} P_{l-1} F_2
\right.
\nonumber \\
& & \left. + \frac{l + 1}{2l (2l+1)} \left( P_{l+1} - P_{l-1} \right) F_3
\right. \nonumber \\
& & \left. + \frac{1}{2 ( 2 l - 1 )} (P_l - P_{l-2} ) F_4 \right] \; , \\
M_{l+} & = & \int_{-1}^{1} dx
\left[ \frac{1}{2 (l + 1)} P_l F_1 - \frac{1}{2 ( l + 1 )} P_{l+1} F_2
\right.
\nonumber \\
& & \left. + \frac{1}{2 (l+1) (2 l + 1)} \left( P_{l+1} - P_{l-1} \right) F_3
\right] \; , \\
M_{l-} & = & \int_{-1}^{1} dx
\left[ - \frac{1}{2 l} P_l F_1 + \frac{1}{2 l} P_{l-1} F_2
\right.
\nonumber \\
& & \left. + \frac{1}{2 l (2l+1)} \left( P_{l-1} - P_{l+1} \right) F_3
\right] \; , \\
L_{l+} & = & \int_{-1}^{1} dx
\frac{1}{2 (l + 1)} \left[ P_l F_5 + P_{l+1} F_6 \right] \; , \\
L_{l-} & = & \int_{-1}^{1} dx
\frac{1}{2 l} \left[ P_l F_5 + P_{l+1} F_6 \right] \; .
\end{eqnarray}
The longitudinal multipole amplitudes $L_{l \pm}$
are related to the scalar amplitudes
 $S_{l \pm}$, which originate from the
time-like component of the transition current (\ref{CGLNcurr}),
by current conservation,
\begin{equation}
\omega S_{l \pm} = \mid \vec q \mid L_{l \pm} \; .
\end{equation}
Furthermore it is convenient to reconstruct the
multipoles from amplitudes $A_{l \pm}$, $B_{l \pm}$ and $C_{l \pm}$, which
are closely related to the electromagnetic resonance couplings
$A_{1/2}^N$, $A_{3/2}^N$ and $C_{1/2}^N$ to be discussed in Sect.
\ref{modelres}:
\begin{eqnarray}
E_{l+} & = & \frac{1}{l + 1} A_{l +} + \frac{l}{2(l + 1)}B_{l+} \;
, \label{MULTBW1}\\
E_{l-} & = & -\frac{1}{l} A_{l -} + \frac{l + 1}{2l}B_{l -} \; , \\
M_{l+} & = & \frac{1}{l + 1} A_{l +} - \frac{l + 2}{2(l + 1)}B_{l+} \; , \\
M_{l-} & = & \frac{1}{l} A_{l -} + \frac{l - 1}{2l}B_{l-} \; , \\
S_{l+} & = & \frac{1}{l + 1} C_{l + } \; , \\
S_{l-} & = & - \frac{1}{l} C_{l -} \; . \label{MULTBW6}
\end{eqnarray}

\section{A Model for Eta Electroproduction}
\label{themodel}
\label{modelres}
The dominant process for eta photoproduction is
given by nucleon isobar excitation. In addition, background
contributions from Born terms and $t$-channel vector meson exchange will be
considered.

The resonance contributions in the direct channel are
parametrized in the usual manner (see e.g.
\cite{DS78} or \cite{CK94})
by means of Breit-Wigner helicity amplitudes
$A_{l \pm}$, $B_{l \pm}$ and $C_{l \pm}$ in the relevant
partial wave. These amplitudes are related to the
multipole amplitudes through relations (\ref{MULTBW1}) to (\ref{MULTBW6}).
In the construction of our model we will use
the following definitions for these amplitudes:
\begin{eqnarray}
A_{l \pm} & = & \pm K A_{1/2}^N, \\
B_{l \pm} & = & \mp K \sqrt{\frac{4}{l ( l + 2 )}} A_{3/2}^N, \\
C_{l \pm} & = & \pm K C_{1/2}^N.
\end{eqnarray}
The factor $K$
\cite{PDG76}
describes the propagation and decay of an
$N^*$ resonance and consists of a Breit-Wigner term and a phase space factor
for the partial wave,
\begin{equation}
\label{BWres}
K = \sqrt{\frac{k_{\gamma}^{cm}}{\mid \vec k \mid}
\frac{m}{W} \frac{ \Gamma_{\eta}}{\pi \left( 2 J + 1 \right) }}
\frac{M^*}{{M^*}^2 - W^2 - i W \Gamma}.
\end{equation}

%
The electromagnetic excitation of a resonance off the nucleon $N$
($N \in \{ p, n \}$)
is described by the electromagnetic
helicity amplitudes $A_{1/2}^N$, $A_{3/2}^N$
and $C_{1/2}^N$
(see e.g. \cite{CKO69} or \cite{War90}),
which consist of a standard kinematical factor
$R = \sqrt{2 \pi \alpha / k_{\gamma}^{cm}} $
and
the matrix elements of spherical components
$ {J^{[j]}_{\lambda}}^{int}$ of the
electromagnetic current operator at the photon-baryon vertex:
\begin{eqnarray}
A_{1/2}^N & = & R \langle N^*, J_z =
\frac{1}{2} \mid
{J^{[1]}_1}^{int} \mid N, J_z = - \frac{1}{2} \rangle, \\
A_{3/2}^N & = & R \langle N^*, J_z =
\frac{3}{2} \mid
{J^{[1]}_1}^{int} \mid N, J_z = \frac{1}{2} \rangle, \\
C_{1/2}^N & = & R \langle N^*, J_z =
\frac{1}{2} \mid
{J^{[0]}_0}^{int} \mid N, J_z = \frac{1}{2} \rangle.
\end{eqnarray}
The simplest choice for deriving these matrix
elements is the non-relativistic quark model, which we will use
in our standard calculations. Within the framework of this model the
current operator
is given by \begin{eqnarray}
\vec J \left( \vec q \right) & = & \frac{1}{2 m_q}
\sum_{j = 1}^{3} \left( e_j
\left\{ \vec p_j, e^{i \vec q \cdot \vec x_j} \right\}
- i \mu_j \vec q \times \vec
\sigma_j e^{i \vec q \cdot \vec x_j} \right), \\
\rho \left( \vec q \right) & = & \sum_{j = 1}^{3} e_j
e^{i \vec q \cdot \vec x_j} \; .
\end{eqnarray}
The charge of a constituent quark $e_j$ is given in units of the
electric charge $e$, and the constituent quark magnetic moment is given
as $\mu_j = e_j$, because we treat the constituent quark as Dirac particles.
For the wave functions of the baryonic states relevant in
eta production we apply standard harmonic oscillator wavefunctions
with configuration mixing in the $S_{11} \left( 1535 \right) $ and
$D_{13} \left( 1520 \right) $ states according to \cite{IK78}. The
$u$- and $d$-quark masses are taken to be $ m_q = 350 \, \, {\mathrm MeV} $
\cite{IK78},
and for the
oscillator spring constant we use the value of $300 \, \, {\mathrm MeV}$
\cite{IK78}. The matrix elements are evaluated in the equal velocity frame.
We will compare the electroproduction
results obtained in this model (to  be called
M1 in the following) with the relativized
constituent quark model (M2) by Warns et al. \cite{War90}, the relativistic
light cone calculation (M3) by Konen and Weber \cite{KW90} and the
light-front approach (M4) by Capstick and Keister \cite{CK94}. Figs.
\ref{cops1a1pn},
\ref{copd1a1pn} and \ref{copp1a1pn}
present the matrix elements $A_{1/2}^N$, $A_{3/2}^N$ and
$C_{1/2}^N$ for the resonances $S_{11}$, $D_{13}$ and $P_{11}$ as a
function of the momentum transfer $Q^2$.

In (\ref{BWres}) the
branching fraction $\Gamma_{\eta} / \Gamma$ for resonance
decay into nucleon and eta meson determines the strength of the coupling of
the eta meson to the nucleon. In our model we use the branching fractions
listed in Table \ref{respartab}.
For all hadronic widths necessary in the description of the
resonance contributions we introduce the energy dependence suggested
in \cite{Ma84} and \cite{MS92},
which results in the correct threshold behaviour of the multipole
amplitudes with respect to the momentum of the outgoing meson
$\vec k$ \cite{Do72},
\begin{eqnarray}
E_{l \pm} \sim \mid \vec k \mid^l, \; \;&
M_{l \pm} \sim \mid \vec k \mid^l, \; \;&
L_{l \pm} \sim \mid \vec k \mid^l \; .
\end{eqnarray}
The energy dependent total width of a resonance $\Gamma(W)$
is given by
the sum over the partial widths of all possible decay channels,
\begin{equation}
\Gamma(W) = \sum_r \Gamma_r (W).
\end{equation}
We construct the energy dependence of $\Gamma_r$ according to the
$\pi N$ scattering analysis in \cite{Ma84}. The experimental data available
on branching fractions usually refer to the branching fraction at the
resonance energy. The energy dependence is introduced by the
function $\rho_j$, which takes into account the penetration of the
angular momentum barrier for total angular momentum $j$,
\begin{equation}
\Gamma_r (W) = \Gamma_r (M^*) \frac{\rho_j (W)}{\rho_j (M^*)}.
\end{equation}
The functions $\rho_j$ are constructed from the Blatt-Weisskopf
functions $B_l$ \cite{BW52},
\begin{equation}
\rho_j(M^*) = \frac{\mid \vec k \mid}{W} B_l^2 ( \mid
\vec k \mid R ).
\end{equation}
The momentum $\vec k$ is the momentum of the outgoing meson in the
c.m. frame. The effective interaction radius $R$ will be fixed at $1.0$ fm.
For the resonances under consideration we need
the Blatt-Weisskopf functions for $l=0,1,2$,
\begin{eqnarray}
B_0 (x) & = & 1, \\
B_1 (x) & = & \frac{x}{\sqrt{1 + x^2}}, \\
B_2 (x) & = & \frac{x^2}{\sqrt{x^4 + 3 x^2 + 9}}.
\end{eqnarray}
If one of the particles in the final state is unstable and decays into
two particles with $m_1$ and $m_2$, while the stable
particle in the final state has mass $m_3$, we calculate $\rho_j ( W )$
by integration over the possible energy range for the
decay products,
\begin{equation}
\rho_j \left( W \right) = \int_{m_1 + m_2}^{W - m_3}
\sigma \left( {\cal{M}} \right) \frac{\mid \vec k \mid}{W} B_l^2 (
\mid \vec k \mid R )
d {\cal{M}},
\end{equation}
where
$\sigma(\cal{M})$ is a Breit-Wigner distribution,
\begin{equation}
\sigma ({\cal{M}}) = \frac{1}{2 \pi}
\frac{\Gamma_0}{({\cal{M}} - M_0)^2 + (\Gamma_0 /2)^2} \; .
\end{equation}
In this parametrization $M_0$ is the mass of the unstable particle and
$\Gamma_0$ its partial width.
In the
construction of the hadronic widths we essentially use resonance
parameters obtained from the $\pi N$-scattering analysis \cite{Ma84}.
We list the parameter set for the resonances
relevant for our model in Table \ref{respartab}.
\begin{table*}[ht]
\renewcommand{\arraystretch}{1.3}    
\caption
{\label{respartab}
\small
Resonance parameters in our model. The symbol $\varepsilon$ represents
an isoscalar two-pion-state with mass and width of $800 \; \; {\mathrm
MeV}$
for the parametrization of uncorrelated two-pion
production.
}
\begin{tabular}{l l l l l}
\hline\noalign{\smallskip}
& $S_{11}(1535)$ & $ P_{11}(1440)  $ & $ D_{13}(1520) $ & $ D_{15}(1675) $
\\ \hline
$ M^* \, \, [{\mathrm MeV}] $ & $1544.0 $ & $1462.0 $
& $1524.0 $ & $ 1676.0 $ \\
$ \Gamma \, \, [{\mathrm MeV}] $ & $166.0 $ & $391.0 $
& $124.0 $ & $ 179.0 $\\
$ A_{1/2}^p \, \, [10^{-3} {\mathrm GeV}^{-1/2}] $
& 107 & $-72$ & $-22$ & 19\\
$ A_{1/2}^n \, \, [10^{-3} {\mathrm GeV}^{-1/2}] $
& $-96$ & 52 & $-62$ & $-47$\\
$ A_{3/2}^p \, \, [10^{-3} {\mathrm GeV}^{-1/2}] $
& - & - & 163 & 19 \\
$ A_{3/2}^n \, \, [10^{-3} {\mathrm GeV}^{-1/2}] $
& - & - & $-137$ & $-69$\\
$ C_{1/2}^p \, \, [10^{-3} {\mathrm GeV}^{-1/2}] $
& 58 & $-52$ & $-93$ & 0\\
$ C_{1/2}^n \, \, [10^{-3} {\mathrm GeV}^{-1/2}] $
& $-72$ & 0 & 99 & 0\\
meson production multipole & $E_{0+}$, $L_{0+}$ & $M_{1-}$, $L_{1-}$
& $E_{2-}$,
$M_{2-}$, $L_{2-}$ & $E_{2+}$, $M_{2+}$, $L_{2+}$ \\
electromagnetic multipole & $E1$, $C1$ & $M1$, $C0$
& $E1$,
$M2$, $C1$ & $E3$, $M2$, $C3$\\
\hline
$\mid f \rangle$ & $\Gamma_f(M^*)/ \Gamma(M^*)$ &
$\Gamma_f(M^*)/ \Gamma(M^*)$ &
$\Gamma_f(M^*)/ \Gamma(M^*)$ &
$\Gamma_f(M^*)/ \Gamma(M^*)$ \\
\hline
$\eta N $ & 0.50 & from \protect{\cite{BT91}} & 0.001 & 0.01\\
$\pi N $ & 0.40 & 0.69 & 0.59 & 0.47 \\
$\varepsilon N $ & 0.10 & 0.09 & - & -\\
$(\pi \Delta)_S $ & - & - & 0.05 & - \\
$(\pi \Delta)_P $ & - & 0.22 & - & - \\
$(\pi \Delta)_D $ & - & - & 0.15 & 0.52 \\
$(\rho_3 N)_D $ & - & - & 0.21 & - \\
\noalign{\smallskip}\hline
\end{tabular}
\end{table*}
Unfortunately, the branching fraction $\Gamma_{\eta} / \Gamma$ of the
Roper resonance cannot
be treated within this parametrization, because the
resonance energy is located below threshold. In this case we use the
parametrization of the dynamical model by Bennhold and Tanabe \cite{BT91}
for the energy dependence
of the partial width $\Gamma_{\eta}$.

For the evaluation of the Feynman diagrams contributing to the non-resonant
background we use effective interaction lagrangian densities. The Born
diagrams are calculated with the lagrangians
\begin{eqnarray}
{\cal{L}}^{PS}_{\eta N N} & = &
- i g_{\eta N N} \bar{\psi} \gamma_5 \psi \Phi_{\eta} \; , \\
{\cal{L}}_{\gamma N N} & = & -e {\bar{\psi}} \left[ \frac{1 + \tau_0}{2}
                         \gamma_{\mu} A^{\mu} \right. \nonumber \\
& & \left. - \left( \frac{\kappa_p +
\kappa_n}{2}
                         + \frac{\kappa_p - \kappa_n}{2} \tau_0 \right)
                         \frac{\sigma_{\mu \nu}}{2 m } \partial^{\nu}
A^{\mu} \right] \psi \; ,
\end{eqnarray}
where $e = \mid e \mid
> 0$.
The symbol $\kappa_N$ denotes the anomalous magnetic moment of the nucleon
($\kappa_p = 1.79$, $\kappa_n = -1.91$)
and $g_{\eta N N}$ is the eta-nucleon coupling constant.
We have chosen the pseudoscalar eta-nucleon coupling, because it fits
the data for angular distributions above threshold \cite{Kr94,Wi93}
much better than the alternative
pseudovector coupling, which has been demonstrated in \cite{TBK94}. In this
reference the value of the coupling constant was found to be
$g_{\eta N N}^2/ 4 \pi = 0.4$, which is quite consistent with
other approaches \cite{GK80,Pi93}. In the electroproduction case we
introduce the usual Dirac and Pauli form factors, $F_1(Q^2)$ and $F_2(Q^2)$,
for the Dirac and Pauli currents, respectively.
The connection with the standard dipole
fit $F(Q^2) = (1 + Q^2/0.71 \, \, {\mathrm GeV}^2)^{-2}$
for the Sachs form factors
is established by
\begin{eqnarray}
F_1^p & = & \frac{1 + \tau \left( 1 + \kappa_p \right)}{1 + \tau} F \; , \\
F_2^p & = & \frac{1}{1 + \tau} F \; , \\
F_1^n & = & \frac{\tau \kappa_n}{1 + \tau}\left( 1 - \frac{1}{1 + 4 \tau}
\right)F \; , \\
F_2^n & = & \frac{1}{1 + \tau} \left( 1 - \frac{\tau}{1 + 4 \tau} \right) F
\; ,
\end{eqnarray}
with $\tau = Q^2 / 4 m^2$.

The effective lagrangians for the vector meson exchange vertices
are given by
\begin{eqnarray}
{\cal{L}}_{\gamma \eta V} & = & \frac{g_{\gamma \eta V}}{m_{\eta}}
           \varepsilon_{\mu \nu \rho \sigma} \partial^{\mu} A^{\nu}
\Phi_{\eta}
           \partial^{\rho} V^{\sigma} \label{vecww},\\
{\cal{L}}_{V N N} & = & {\bar{\psi}} \left( g_{V_1} \gamma_{\mu}
                        + \frac{g_{V_2}}{2 m} \sigma_{\mu \nu}
\partial^{\nu} \right)
                        V^{\mu} \psi.
\end{eqnarray}
The parameters for the $\rho$ and $\omega$ mesons are listed in Table
\ref{vecmestab}.
\begin{table}[ht]
\renewcommand{\arraystretch}{1.3}    
\caption{
\small
\label{vecmestab}
Parameters for the vector mesons.}
\begin{tabular}{l l l l l l}
\hline\noalign{\smallskip}
$ V $ & $ m_V [{\mathrm MeV}] $ & $ g_{V_1}^2/4 \pi $
& $ g_{V_2} / g_{V_1} $ & $
\Lambda_V [{\mathrm MeV}] $ & $ \lambda_V $ \\
\hline
$ \omega $ & $ 782.6 $ & $ 23 $ & $ 0 $ & $ 1400 $ & $ 0.192 $ \\
$ \rho $ & $ 769.0 $ & $ 0.5 $ & $ 6.1 $ & $ 1800 $ & $ 0.89 $ \\
\noalign{\smallskip}\hline
\end{tabular}
\end{table}
The electromagnetic couplings of the vector mesons $g_{\gamma \eta V}
= e \lambda_V F_V(\vec k_V^2 )$ are determined from the radiative widths.
Following \cite{La88} the form factor $F_V$ is supposed
to have the usual dipole behaviour. The hadronic couplings $g_{V_1}$ and
$g_{V_2}$ are taken from a nuclear potential model \cite{BM90} with the
hadronic dipole form factor
\begin{equation}
F_V^{had} ( \vec k_V^2 ) = \frac{\left(\Lambda_V^2 - m_V^2\right)^2}
{\left( \Lambda_V^2 + \vec k_V^2 \right)^2}.
\end{equation}
At tree level these contributions are real, they have been added to the
real part of the resonance
contribution.
In general this procedure violates
unitarity of the $S$-matrix. However, the unitarity
corrections for most of the multipoles are small or even negligible
in the threshold region, because the background contributions of most
multipoles in eta production are
small compared to the resonance contributions. Unfortunately, in
eta production there is no simple
constraint to the phase of the multipoles as
in pion production, because the Watson theorem \cite{Wa54}
is only valid in the
elastic regime, whereas in eta production
there are at least three hadronic reaction channels,
$| N \pi \rangle $,
$| N \eta \rangle $, and
$| N \pi \pi \rangle$.
In particular a consistent treatment of
the state $|N \pi \pi \rangle$
is difficult, because it contains both uncorrelated two pion production and
resonance mechanisms.

\section{Results for Photoproduction}
%
%
The starting point of our model will be
the latest TAPS \cite{Kr94} and Bonn \cite{Wi93} data
on the total cross
section for photoproduction of eta mesons on the proton
\footnote{The Bonn data \protect{\cite{Wi93}} are electroproduction
data at very low momentum transfer, $Q^2 = 0.056 \,
{\mathrm GeV}^2$. Comparison with
Fig. \protect{\ref{sl11}} justifies our treatment of these data as
photoproduction data.}. A Breit-Wigner fit to these data,
\begin{equation}
\sigma_{tot} = \frac{\mid \vec k \mid}{\mid \vec q \mid}
\frac{D {M^*}^2 \Gamma(W)^2}{\left( {M^*}^2 - W^2 \right)^2 + {M^*}^2
\Gamma^2 ( W )} \; ,
\end{equation}
with a momentum
dependent hadronic width,
\begin{equation}
\Gamma(W) = \Gamma(M^*) \left( 0.50
\frac{\mid \vec k \mid}{\mid \vec k^* \mid} +
0.40 \frac{\mid \vec k_{\pi} \mid}{\mid \vec k_{\pi}^* \mid}
+ 0.10 \right),
\end{equation}
results in the following values for the three fit parameters:
$
M^* = 1544.0 \; {\mathrm MeV}, \; \Gamma(M^*) = 166.0 \; {\mathrm MeV},
\; D = 39.0 \; \mu {\mathrm b} \; . $
Note that we use the branching fractions
$\Gamma_{\eta}(M^*) / \Gamma(M^*) = 0.50, \; \;$
$\Gamma_{\pi}(M^*) / \Gamma(M^*) = 0.40$ and
$\Gamma_{\pi \pi}(M^*) / \Gamma(M^*) = 0.10$ as input.
At this point we have assumed,
that the total cross section in this energy region is
completely dominated by the $S_{11} (1535)$, which is justified
by the small background,
$P_{11}$, and $D_{13}$
contributions (see Fig. \ref{photowqinclp}). These additional
contributions modify the shape of the total cross section in
the $S_{11}(1535)$ region only slightly.
Hence the
electromagnetic helicity coupling $A_{1/2}^p$ of the $S_{11}(1535)$
at resonance energy $W = M^*$
is given by
\begin{equation}
\mid A_{1/2}^p \mid = \sqrt{\frac{\mid \vec k \mid}{k_{\gamma}^{cm}}
\frac{M^*}{m} \frac{2 D}{\Gamma_{\eta}(M^*)}} \frac{\Gamma(M^*)}{2}.
\end{equation}
The result of
$107 \times 10^{-3} \; {\mathrm GeV}^{-1/2}$
is significantly larger than the
standard value of
$(68 \pm 10)
\times 10^{-3} \; {\mathrm GeV}^{-1/2}$ in \cite{PDG94}, but
seems to be consistent with the value of
$(95 \pm 11) \times 10^{-3} \; {\mathrm GeV}^{-1/2}$
from the
analysis of eta production data \cite{BM91}. The TAPS analysis \cite{Kr94}
yields values between $110$ and $140 \times
10^{-3} \; {\mathrm GeV}^{-1/2} $
depending on the branching ratios of the $S_{11}$.
Towards higher energies, the background contributions become more important,
and our standard calculation is still consistent with
eta production data from previous experiments \cite{ABB68,CBC68}
within the large error bars.
The latest preliminary data from Bonn \cite{Br94},
however, are in good agreement with our standard calculation in
the energy range up to $W = 1800 \, \, {\mathrm MeV}$.

The TAPS analysis of eta production data for total cross sections
on the deuteron suggests that the elementary cross section
on the neutron is about $80\%$ of the proton value at the resonance peak
\cite{Krprivat},
which corresponds to an electromagnetic helicity coupling of
$A_{1/2}^n = -96 \times 10^{-3} \; {\mathrm GeV}^{-1/2}$
for the $S_{11} (1535)$ under the assumption of resonance dominance
and is again significantly larger than the standard value \cite{PDG92} of
$A_{1/2}^n = (-59 \pm 22) \times 10^{-3} \; {\mathrm GeV}^{-1/2}$.
%
%
The $S_{11}$ dominance assumption is also supported by the rather flat
angular distribution of the unpolarized differential cross section on the
proton
measured by the TAPS collaboration \cite{Kr94} (see Fig. \ref{ds752}).
Our standard calculation with background and resonance
contributions of $S_{11} (1535)$, $D_{13} (1520)$, $D_{15}(1675)$
as well as $P_{11} (1440)$
is consistent with the data except for large angles.
Omission of the Roper resonance results in an
over-estimation of the differential cross section at small angles, whereas
omission of the $D_{13}(1520)$ overestimates the differential cross
section at medium and large angles. As the branching fraction of the
$D_{13}(1520)$ resonance into the eta meson is not known precisely, we
have also investigated a three- and tenfold
increase in $\Gamma_{\eta}$. While the threefold increase in
$\Gamma_{\eta}$ seems to improve the description of the data slightly,
the tenfold increase
is definitely
not consistent with experiment. However, it has to be mentioned that the
strength of a resonance contribution in a given multipole
is determined by the factor
$A_{\lambda}^N \sqrt{\Gamma_{\eta}}/ \Gamma$, where $\lambda$ is the
helicity of the intermediate resonance state. Therefore
the uncertainties in the
photocouplings $A_{\lambda}^N$  do also enter the
strength of the respective multipole.
Fig. \ref{photowqincln}
presents the unpolarized differential eta photoproduction
cross sections off the proton and the neutron as a function of excitation
energy $\nu$ and scattering angle $\Theta_{\eta}$. The calculations
were performed in our standard parameter set, the cross section on the
neutron was normalized to $80\%$ of the proton value at resonance maximum.
It is remarkable that the cross section for the neutron exposes a slight
minimum at $90^o$, whereas the proton cross section has a slight
maximum in this range of $\Theta_{\eta}$. This observation is compatible
with the analysis in \cite{TBK94}.
%
Having fixed the parameters of our model by
unpolarized cross section
data, we now proceed to study the polarization observables for single
and double polarization experiments with beam and
target polarization. All of these observables are accessible at the
laser-backscattering facility GRAAL, where an extensive
experimental program for eta photoproduction is already under way.
Analysing the angular distributions of the
observables in our model, it turns out
that the most interesting effects can be seen
at a scattering angle of $90^o$. For this reason we choose this angle for
the presentation of the excitation functions in Fig. \ref{obs1p}.
The shape of the differential cross section does not expose
any significant dependence on the $D_{13}$ or $P_{11}$ resonance. However,
the non-resonant background destructively
interferes with the $S_{11}$ contribution towards higher photon energies,
which leads to a slight
reduction of the total cross section. The photon asymmetry $\Sigma$ shows a
very characteristic dependence on the contribution of the $D_{13}(1520)$
resonance due to interference of the $E_{2-}$ and $M_{2-}$ multipoles with
the $E_{0+}$ multipole in ${ }^c \! R_{TT}^{00}$.
Although the $D_{13}$ resonance has only a very small branching ratio
into the eta meson, its absence would lead to an almost vanishing
photon asymmetry within a wide range of excitation energies. Switching off
the non-resonant background in the target or recoil asymmetry results in a
value for these observables which is consistent with zero. In the threshold
region the influence of the Roper resonance on the recoil asymmetry might
also be visible in a precision experiment. Inspecting the double polarization
observables for beam and target polarization, it becomes evident that the
observable $-H$ behaves similar to the recoil asymmetry $P$ in the
threshold region. Background and $D_{13}$ effects are also visible in the
$G$ observable above threshold. The $E$ observable, however, is consistent
with unity above threshold in each of the scenarios investigated.
Since there are no
additional spectacular effects in the
double polarization observables for beam and recoil respectively target and
recoil experiments, we will not present our results for these
observables.

\section{Results for Electroproduction}

At the electron scattering facilities ELSA and
CEBAF (as well as at MAMI at low $Q^2$ in the threshold region)
eta electroproduction on the proton can be studied with
very high precision.
As the available data are very scarce and have
large error bars, the new experiments could tremendously extend our
knowledge on this process. To our knowledge
there has been no theoretical investigation of the
eta electroproduction process in the
literature before. Therefore, we will present some observables
relevant for future experiments. Comparing Tables \ref{elprodtab} and
\ref{phprodtab}, it becomes
obvious that the electroproduction process has a significantly richer
phenomenology than the photoproduction process due to the additional
longitudinal component of the photon and the $Q^2$-dependence of all
observables. In our model we will fix the contribution
of longitudinal photons to the excitation of the $S_{11}$ resonance by
the data of an old Bonn experiment \cite{Ni78}, which, of course, has large
error bars. The $Q^2$-dependence of the resonant part of our
electroproduction operator as well as possible longitudinal
contributions of other resonances will be treated within the framework of
quark models. The non-resonant part of our operator already incorporates
the longitudinal components of the transition current $\vec J$, and the
$Q^2$-behaviour is generated by standard dipole form factors.

%
Fig. \ref{sigmal} presents the longitudinal cross section $\sigma_L$ for
kinematics 1
in Table \ref{elkin} in different models, which are more or
less consistent with the existing data point.
In our standard calculations we
will use the non-relativistic constituent quark model with
the amplitude of the resonant part of the $L_{0+}$ multipole normalized
to the data point from \cite{Ni78}.
The reason for the large experimental error bar is due to the fact
that the longitudinal excitation of the $S_{11}(1535)$
is weaker than
the transverse one. This can be clearly seen in Fig. \ref{sl11}.
The transverse/longitudinal separation
of the inclusive cross section,
$\sigma_{tot} = \sigma_T + \varepsilon_L \sigma_L$, leads to a
longitudinal cross section $\varepsilon_L \sigma_L$, which is smaller than
the transverse $\sigma_T$ by more than an order of magnitude. The
data point for the transverse
cross section, however, is in good agreement with our model prediction.
In this article we will only discuss angular distributions of cross
sections that do not require any electron or recoil polarization measurement.
The
complete formula for the differential cross
section describing this scenario is given by (\ref{dsigmafull}).
For kinematics 1 and 2 in Figs. \ref{resp1p}, \ref{resp2p} and
\ref{resp3p}
we present the results for the cross sections
\begin{eqnarray}
\label{dwq}
& &\frac{d \sigma_T^{00}}{d \Omega_{\eta}} = \rho R_T^{00}  \; ,
\frac{d \sigma_L^{00}}{d \Omega_{\eta}} = \rho R_L^{00} \; , \nonumber \\
& & \frac{d \sigma_{TL}^{00}}{d \Omega_{\eta}} = \rho { }^c \! R_{TL}^{00}
\; ,
\frac{d \sigma_{TT}^{00}}{d \Omega_{\eta}} = \rho { }^c \! R_{TT}^{00}
\; ,\\
& & \frac{d \sigma_{TT}^{0x}}{d \Omega_{\eta}} = \rho { }^s \! R_{TT}^{0x}
\; ,
\frac{d \sigma_{TT}^{0y}}{d \Omega_{\eta}} = \rho { }^c \! R_{TT}^{0y}
\; ,\nonumber
\end{eqnarray}
where $\rho = \mid \vec k \mid /
k_{\gamma}^{cm}$.

The resulting transverse cross section $d \sigma_T^{00} /d \Omega$
has a rather flat
angular distribution as in the
photoproduction case. The longitudinal cross section
$d \sigma_L^{00} / d \Omega$ is smaller than the transverse by about one
order
of magnitude and has a maximum around $100^o$
due to the $D_{13}(1520)$ contribution. Of course, the magnitude of this cross
section depends strongly on the prediction of the different quark models for
$C_{1/2}^p$. The cross sections $d \sigma_{TT}^{00} / d \Omega$ and
$d \sigma_{TL}^{00} / d \Omega$ are very sensitive to the presence of the
resonance $D_{13}(1520)$. The transverse-transverse interference cross
section almost vanishes without this resonance, and the shape of
the transverse-longitudinal interference cross section changes completely
when the resonance is decoupled.
Finally we will have a look at the cross sections belonging to the response
functions
${ }^c \! R_{TT}^{0y}$ and
${ }^s \! R_{TT}^{0x}$, which
correspond to the photoproduction observables $- P$ and $H$.
In electroproduction they could be determined in an experiment with a
polarized target or a recoil polarimeter. From
photoproduction we already know the sensitivity of these observables
to background and Roper resonance contributions.
At larger momentum transfer, $Q^2 = 0.120 \; {\mathrm GeV}^2,$ we observe
different signs for
the calculations with and without Roper resonance.
An experiment at
this kinematics would be invaluable in order to
detect a signature of the Roper resonance in
eta production.
Unfortunately, there are large discrepancies between the different quark
models concerning the electromagnetic couplings $A_{1/2}^p$ and $C_{1/2}^p$
of the Roper resonance.
In addition to this, the Roper
resonance value for
$A_{1/2}^p$ in \cite{PDG92} is not fixed very precisely by experiment as
well.
\begin{table}[ht]
\renewcommand{\arraystretch}{1.3}
\caption{\small \label{elkin}Kinematics
investigated in electroproduction.}
\begin{tabular}{l l l}
\hline\noalign{\smallskip}
 & Kinematics 1 & Kinematics 2 \\ \hline
$ \varepsilon_i \left[ {\mathrm MeV} \right] $ & $ 1337.5 $ & $ 989.4 $ \\
$ \varepsilon_f \left[ {\mathrm MeV} \right] $ & $ 344.5 $ & $ 142.2 $ \\
$ \Theta_e \left[^o \right] $ & $ 55.0 $ & $ 55.0 $ \\
$ W \left[ {\mathrm MeV} \right] $ & $ 1533.0 $ & $ 1533.0 $ \\
$ Q^2 \left[ {\mathrm GeV}^2 \right] $ & $ 0.393 $ & $ 0.120 $ \\
$ \varepsilon $ & $ 0.345 $ & $ 0.209 $ \\
\noalign{\smallskip}\hline
\end{tabular}
\end{table}

\section{Conclusion}
Photo- and electroproduction of eta mesons are fastly developing
fields in intermediate energy physics. Due to the isospin singlet
nature, the eta meson can supply valuable
information on the structure of the
nucleon complementary to that obtained from pion production. Such
experiments are under investigation
at various new facilities. The TAPS collaboration has presented
results on inclusive and differential photoproduction
cross sections off the proton,
which were obtained at the MAMI facility, and
at the ELSA facility
the electroproduction and photoproduction experiments at higher energies
are being analysed.
Further experiments at these sites are under way with the aim to
measure cross sections off the deuteron as
well as polarization observables off the proton and the deuteron.
In the future, CEBAF will join this experimental field with
$\eta$ and $\eta'$ electroproduction, investigating
various resonance contributions to the process. Last but not the least, the
laser-backscattering facility GRAAL is an excellent tool for
the investigation of polarization degrees of freedom in
eta photoproduction.

In this article we have studied the complete set of response functions for
photo-
and electroproduction of the eta meson including
 beam, target and recoil polarization degrees of freedom.
Since future experiments will have to be analysed in terms of multipoles,
we have also presented the multipole decomposition of the response functions.
Near threshold eta production is dominated by the
decay of the nucleon resonance $S_{11}(1535)$, which contributes
to the $E_{0+}$ and $L_{0+}$ multipoles.
We derived the $E_{0+}$ multipole from the latest Mainz \cite{Kr94}
and Bonn \cite{Wi93} data for the total photoproduction cross section.
The $L_{0+}$ multipole was reconstructed from an old Bonn experiment
\cite{Ni78}, the $Q^2$-dependence of this multipole was investigated within
the framework of different quark models \cite{CK94,CKO69,War90,KW90}.
The transition current contained the non-resonant background from the model
presented in \cite{TBK94} as well as phenomenological resonance
contributions.
Having fixed the dominating $S$-wave multipoles, we were able to
investigate additional resonance contributions of the $P_{11}(1440)$ and
$D_{13}(1520)$ resonances. These resonances generate only weak contributions
in the unpolarized cross sections. However, they produce tremendous
effects in some of the polarization observables by interference with the
dominant $S_{11}(1535)$ multipole. We estimated the strength of the
Roper and $D_{13}$-resonance multipoles by means of a standard
Breit-Wigner approach, having determined the electromagnetic
resonance couplings by the helicity elements
$A_{1/2}^N$,
$A_{3/2}^N$ and
$C_{1/2}^N$, which can be calculated within the framework of quark
models. The resonance couplings to the eta meson are
determined by the branching fraction $\Gamma_{\eta} (W)/ \Gamma(W)$ in our
phenomenological model.
Our estimate of these additional resonance contributions is compatible
with the angular distributions of differential
cross sections for photoproduction off the proton
measured by the TAPS collaboration. However, it turns out
that the measurement of polarization observables can give
more reliable information on the resonance contributions. In
photoproduction the photon asymmetry should yield precise
constraints on the coupling of the $D_{13}(1520)$ to
the eta meson. A possible Roper contribution significantly
changes the shape of the
photoproduction observables $P$ and $H$. These observables are also
sensitive to the non-resonant background and
provide an additional test of the $\eta N N$--coupling constant
\cite{TBK94}. With regard to photoproduction off the neutron,
it would be interesting to see
whether the
saddle structure in the differential cross section leads to any effects
possibly visible in an experiment on the deuteron.

Concerning the electroproduction of
eta mesons, our calculations suffer from the large
error bars
of the existing longitudinal cross sections off the proton
used as input.
Nevertheless, eta electroproduction
is a promising field, because the determination of the
$L_{0+}$ multipole from a transverse/longitudinal
separation of the total cross section provides a good test of the
quark models describing the coupling of a longitudinal
photon to the $S_{11}(1535)$ resonance. Furthermore, the
separation of the differential cross section $d \sigma_{v} / d \Omega_{\eta}$
into transverse, longitudinal, transverse-transverse and
transverse-longitudinal parts can provide additional interesting
information on the mechanisms present in eta production. The
transverse-transverse interference cross section
$d \sigma_{TT}^{00} / d \Omega_{\eta}$ is related to the observable
$\Sigma$ in photoproduction.
For this reason it also has a characteristic dependence
on the $D_{13}$-contribution.
The presence
of a $D_{13}$-contribution also changes the shape of
$d \sigma_{TL}^{00} / d \Omega_{\eta}$ drastically. The Roper and background
contributions can be seen best in the response functions ${ }^s \!
R_{TT}^{0x}$ and
${ }^c \! R_{TT}^{0y}$, whose
determination, however,
requires a target (or recoil) polarization experiment.

In conclusion eta production exposes
a rich phenomenology, which can significantly
enlarge our knowledge on the structure of the nucleon. We
hope
the material we have presented will be useful
for the experimental projects under way or even
stimulate new experiments to determine some
of the polarization observables discussed in this contribution.

\appendix
\section{Identical Response Functions}
\label{identresp}
A response function denoted with a star in Table \ref{elprodtab} is
identical to another response function via one of the following equations:
\begin{eqnarray}
& & R_{T}^{00} = - { }^c \! R_{TT}^{y' y},
 R_{T}^{0y} = - { }^c \! R_{TT}^{y' 0},
 R_{T}^{y'0} = - { }^c \! R_{TT}^{0 y}, \nonumber \\
& & R_{T}^{x'x} = - { }^c \! R_{TT}^{z'z},
 R_{T}^{x'z} = { }^c \! R_{TT}^{z'x},
 R_{T}^{z'x} = { }^c \! R_{TT}^{x'z}, \nonumber \\
& & R_{T}^{z'z} = - { }^c \! R_{TT}^{x'x},
 R_{L}^{00} = - R_{L}^{y'y},
 R_{L}^{0y} = - R_{L}^{y'0}, \nonumber \\
& & R_{L}^{x'x} = - R_{L}^{z'z},
 R_{L}^{x'z} = R_{L}^{z'x},
 { }^c \! R_{TL}^{00} = -{ }^c \! R_{TL}^{y'y}, \nonumber \\
& & { }^c \! R_{TL}^{0y} = -{ }^c \! R_{TL}^{y'0},
 { }^c \! R_{TL}^{x'x} = -{ }^c \! R_{TL}^{z'z},
 { }^c \! R_{TL}^{z'x} = { }^c \! R_{TL}^{x'z}, \nonumber \\
& & { }^s \! R_{TL}^{0x} = -{ }^c \! R_{TL'}^{y'z},
 { }^s \! R_{TL}^{0z} = -{ }^c \! R_{TL'}^{y'x},
 { }^s \! R_{TL}^{x'0} = -{ }^c \! R_{TL'}^{z'y}, \nonumber \\
& & { }^s \! R_{TL}^{z'0} = { }^c \! R_{TL'}^{x'y},
 { }^c \! R_{TT}^{00} = - R_{T}^{y'y},
 { }^s \! R_{TT}^{0x} = R_{TT'}^{y'z}, \nonumber \\
& & { }^s \! R_{TT}^{z'0} = R_{TT'}^{x'y},
 { }^s \! R_{TT}^{0z} = -R_{TT'}^{y'x},
 { }^s \! R_{TT}^{x'0} = -R_{TT'}^{z'y}, \nonumber \\
& & { }^c \! R_{TL'}^{0x} = -{ }^s \! R_{TL}^{y'z},
 { }^c \! R_{TL'}^{0z} = { }^s \! R_{TL}^{y'x},
 { }^c \! R_{TL'}^{x'0} = { }^s \! R_{TL}^{z'y}, \nonumber \\
& & { }^c \! R_{TL'}^{z'0} = -{ }^s \! R_{TL}^{x'y},
 { }^s \! R_{TL'}^{00} = -{ }^s \! R_{TL'}^{y'y},
 { }^s \! R_{TL'}^{0y} = -{ }^s \! R_{TL'}^{y'0},\nonumber \\
& & { }^s \! R_{TL'}^{x'x} = -{ }^s \! R_{TL'}^{z'z},
 { }^s \! R_{TL'}^{z'x} = { }^s \! R_{TL'}^{x'z},
 R_{TT'}^{0x} = -{ }^s \! R_{TT}^{y'z}, \nonumber \\
& & R_{TT'}^{0z} = { }^s \! R_{TT}^{y'x},
 R_{TT'}^{x'0} = { }^s \! R_{TT}^{z'y},
 R_{TT'}^{z'0} = -{ }^s \! R_{TT}^{x'y}. \nonumber
\end{eqnarray}
\section{Polarization Observables and Response Functions}
\label{photoelektro}
The polarization observables in photoproduction can be expressed in terms
of response functions via the following relations:
\begin{eqnarray}
& & d \sigma / d \Omega = \rho R_T^{00} , \; \Sigma = - { }^c \! R_{TT}^{00} /
R_{T}^{00}, \nonumber \\
& & T = R_{T}^{0y} / R_{T}^{00} , \;
P = R_{T}^{y'0} / R_{T}^{00} ,
\nonumber \\
& & E = - R_{TT'}^{0z} / R_{T}^{00}, \;
F = R_{TT'}^{0x} / R_{T}^{00}, \nonumber \\
& & G = - { }^s \! R_{TT}^{0z} / R_{T}^{00}, \;
H = { }^s \! R_{TT}^{0x} / R_{T}^{00},
\nonumber \\
& & O_{x'} = { }^s \! R_{TT}^{x'0} / R_{T}^{00}, \;
O_{z'} = { }^s \! R_{TT}^{z'0} / R_{T}^{00}, \nonumber \\
& & C_{x'} = - R_{TT'}^{x'0} / R_{T}^{00}, \;
C_{z'} = - R_{TT'}^{z'0} / R_{T}^{00},
\nonumber \\
& & T_{x'} = R_{T}^{x'x} / R_{T}^{00}, \;
T_{z'} = R_{T}^{z'x} / R_{T}^{00}, \nonumber \\
& & L_{x'} = - R_{T}^{x'z} / R_{T}^{00}, \;
L_{z'} = R_{T}^{z'z} / R_{T}^{00}.
\nonumber
\end{eqnarray}
\section{Response Functions and CGLN Amplitudes}
\label{CGLNapp}
\begin{eqnarray}
R_{T}^{00} & = & \mid F_1 \mid^2 + \mid F_2 \mid^2
+ \frac{\sin^2 \Theta}{2} \left(
\mid F_3 \mid^2 + \mid F_4 \mid^2 \right)
\nonumber \\
& &
+ \Re \{ \sin^2 \Theta (F_2^* F_3 + F_1^* F_4 + \cos \Theta F_3^* F_4
) \nonumber \\
& & - 2 \cos \Theta F_1^* F_2 \},
\nonumber
\\
R_{T}^{0y} & = & \Im \left\{ \sin \Theta \left( F_1^* F_3 - F_2^* F_4
+ \cos \Theta (F_1^* F_4 - F_2^* F_3) \right. \right. \nonumber \\
& & \left. \left. - \sin^2 \Theta F_3^* F_4
\right) \right\},
\nonumber
\\
R_{T}^{y'0} & = & \Im \left\{ \sin \Theta \left( - 2 F_1^* F_2 - F_1^* F_3
+ F_2^* F_4  \right. \right.
\nonumber \\
& & \left. \left. + \cos \Theta \left( F_2^* F_3 - F_1^* F_4 \right)
+ \sin^2 \Theta F_3^* F_4
\right) \right\},
\nonumber
\\
R_{T}^{x' x} & = & \Re \{ \sin^2 \Theta (
- F_1^* F_3 - F_2^* F_4
- F_3^* F_4
 \nonumber \\
& & - \frac{1}{2} \cos \Theta \left(
\mid F_3 \mid^2 + \mid F_4 \mid^2
\right) ) \},
\nonumber
\\
R_{T}^{x' z} & = & \{ \sin \Theta
(
\mid F_1 \mid^2 - \mid F_2 \mid^2
+ \frac{1}{2} \sin^2 \Theta
(
\mid F_4 \mid^2 - \mid F_3 \mid^2
)
\nonumber \\
& &
- F_2^* F_3 + F_1^* F_4
+ \cos \Theta \left( F_1^* F_3 - F_2^* F_4 \right)
) \},
\nonumber
\\
R_{T}^{z' x} & = & \Re \{ \sin \Theta
( - F_2^* F_3 + F_1^* F_4
+ \cos \Theta \left( F_1^* F_3 - F_2^* F_4 \right)
\nonumber \\
& & + \frac{1}{2} \sin^2 \Theta
( \mid F_4 \mid^2 - \mid F_3 \mid^2
) ) \},
\nonumber
\\
R_{T}^{z' z} & = & \Re \{
2 F_1^* F_2
- \cos \Theta \left(
\mid F_1 \mid^2 + \mid F_2 \mid^2 \right)
 \nonumber \\
& & + \sin^2 \Theta
\left(
F_1^* F_3 + F_2^* F_4 + F_3^* F_4 \right)  \nonumber \\
& & + \frac{1}{2} \cos \Theta \sin^2 \Theta (
\mid F_3 \mid^2 + \mid F_4 \mid^2
)
\},
\nonumber
\\
R_{L}^{00} & = & \Re \left\{
\mid F_5 \mid^2 + \mid F_6 \mid^2 + 2 \cos \Theta
F_5^* F_6 \right\},
\nonumber
\\
R_{L}^{0y} & = & - 2 \sin \Theta \Im \left\{ F_5^* F_6 \right\},
\nonumber
\\
R_{L}^{x' x} & = & \Re \left\{ - 2 F_5^* F_6
- \cos \Theta \left( \mid F_5 \mid^2 + \mid F_6 \mid^2 \right) \right\},
\nonumber
\\
R_{L}^{z' x} & = & \sin \Theta \left( \mid F_6 \mid^2 - \mid F_5 \mid^2
\right),
\nonumber
\\
{ }^c \! R_{TL}^{00} & = & \sin \Theta \Re
\left\{ - F_2^* F_5 - F_3^* F_5 - F_1^* F_6 - F_4^* F_6
\right. \nonumber \\
& & \left. - \cos \Theta \left( F_4^* F_5 + F_3^* F_6 \right)  \right\},
\nonumber
\\
{ }^s \! R_{TL}^{0x} & = & \Im \left\{
- F_1^* F_5 + F_2^* F_6 + \cos \Theta \left(
F_2^* F_5 - F_1^* F_6 \right)
\right\},
\nonumber
\\
{ }^c \! R_{TL}^{0y} & = & \Im \left\{
- F_1^* F_5 + F_2^* F_6 + \cos \Theta \left( F_2^* F_5 - F_1^* F_6 \right)
\right. \nonumber \\
 & &
\left. + \sin^2 \Theta \left( F_3^* F_6 - F_4^* F_5 \right) \right\},
\nonumber
\\
{ }^s \! R_{TL}^{0z} & = & \sin
\Theta \Im \left\{ F_2^* F_5 + F_1^* F_6 \right\},
\nonumber
\\
{ }^s \! R_{TL}^{x'0} & = & \Im \left\{ - F_2^* F_5 + F_1^* F_6 +
\cos \Theta \left( F_1^* F_5 - F_2^* F_6 \right) \right\},
\nonumber
\\
{ }^s \! R_{TL}^{z'0} & = & \sin
\Theta \Im \left\{ F_1^* F_5 + F_2^* F_6 \right\},
\nonumber
\\
{ }^c \! R_{TL}^{x' x} & = & \sin \Theta \Re \left\{ F_1^* F_5 + F_4^* F_5 +
F_2^* F_6 + F_3^* F_6 \right. \nonumber \\
& & \left. + \cos \Theta \left( F_3^* F_5 + F_4^* F_6
\right) \right\},
\nonumber
\\
{ }^c \! R_{TL}^{z' x} & = & \Re \left\{
F_2^* F_5 - F_1^* F_6 + \cos \Theta \left( F_2^* F_6 - F_1^* F_5 \right)
\right. \nonumber \\
& & \left. + \sin^2 \Theta \left( F_3^* F_5 - F_4^* F_6 \right) \right\},
\nonumber
\\
{ }^c \! R_{TT}^{00} & = & \frac{1}{2} \sin^2 \Theta \left\{
\mid F_3 \mid^2 + \mid F_4 \mid^2 \right\}
\nonumber \\
& & + \sin^2 \Theta \Re \left\{
F_2^* F_3 + F_1^* F_4 + \cos \Theta F_3^* F_4 \right\},
\nonumber
\\
{ }^s \! R_{TT}^{0x} & = & \sin \Theta \Im \left\{
2 F_1^* F_2 + F_1^* F_3 - F_2^* F_4 \right. \nonumber \\
& & \left. + \cos \Theta
\left( F_1^* F_4 - F_2^* F_3 \right) \right\},
\nonumber
\\
{ }^s \! R_{TT}^{0z} & = & - \sin^2 \Theta \Im \left\{
F_2^* F_3 + F_1^* F_4 \right\},
\nonumber
\\
{ }^s \! R_{TT}^{x'0} & = & \sin \Theta \Im \left\{ F_2^* F_3 - F_1^* F_4
+ \cos \Theta \left( F_2^* F_4 - F_1^* F_3 \right) \right\},
\nonumber
\\
{ }^s \! R_{TT}^{z'0} & = & - \sin^2 \Theta \Im
\left\{ F_1^* F_3 + F_2^* F_4 \right\},
\nonumber
\\
{ }^s \! R_{TL'}^{00} & = & - \sin \Theta \Im \left\{
F_2^* F_5 + F_3^* F_5 + F_1^* F_6 + F_4^* F_6 \right. \nonumber \\
& & \left. + \cos \Theta \left(
F_4^* F_5 + F_3^* F_6 \right) \right\},
\nonumber
\\
{ }^c \! R_{TL'}^{0x} & = & \Re \left\{ - F_1^* F_5 + F_2^* F_6 + \cos \Theta
\left( F_2^* F_5 - F_1^* F_6 \right) \right\},
\nonumber
\\
{ }^s \! R_{TL'}^{0y} & = & \Re \left\{ F_1^* F_5 - F_2^* F_6
+ \cos \Theta \left( F_1^* F_6 - F_2^* F_5 \right)
\right. \nonumber \\
& & \left. + \sin^2 \Theta \left( F_4^* F_5 - F_3^* F_6 \right) \right\},
\nonumber
\\
{ }^c \! R_{TL'}^{0z} & = & \sin \Theta \Re
\left\{ F_2^* F_5 + F_1^* F_6 \right\},
\nonumber
\\
{ }^c \! R_{TL'}^{x'0} & = & \Re \left\{ - F_2^* F_5 + F_1^* F_6
+ \cos \Theta \left( F_1^* F_5 - F_2^* F_6 \right) \right\},
\nonumber
\\
{ }^c \! R_{TL'}^{z'0} & = & \sin \Theta \Re \left\{ F_1^* F_5 +
F_2^* F_6 \right\},
\nonumber
\\
{ }^s \! R_{TL'}^{x' x} & = & \sin \Theta \Im \left\{
F_1^* F_5 + F_4^* F_5 + F_2^* F_6 + F_3^* F_6
\right. \nonumber \\
& & \left. + \cos \Theta \left( F_3^* F_5 + F_4^* F_6 \right) \right\},
\nonumber
\\
{ }^s \! R_{TL'}^{z' x} & = & \Im \left\{ F_2^* F_5 - F_1^* F_6
+ \cos \Theta \left( - F_1^* F_5 + F_2^* F_6 \right)
\right. \nonumber \\
& & \left. + \sin^2 \Theta \left( F_3^* F_5 - F_4^* F_6 \right) \right\},
\nonumber \\
R_{TT'}^{0x} & = & \sin \Theta \Re \left\{
F_1^* F_3 - F_2^* F_4 + \cos \Theta \left( - F_2^* F_3 + F_1^* F_4 \right)
\right\},
\nonumber
\\
R_{TT'}^{0z} & = & - \mid F_1 \mid^2 - \mid F_2 \mid^2
\nonumber \\
& & + \Re \left\{ 2 \cos \Theta F_1^* F_2 - \sin^2 \Theta
\left( F_2^* F_3 + F_1^* F_4 \right) \right\},
\nonumber
\\
R_{TT'}^{x'0} & = & \sin \Theta \Re \left\{
- \mid F_1 \mid^2 + \mid F_2 \mid^2 +
F_2^* F_3
\right. \nonumber \\
& & \left. - F_1^* F_4 + \cos \Theta ( F_2^* F_4 - F_1^* F_3 )
\right\},
\nonumber
\\
R_{TT'}^{z'0} & = & \Re \left\{ - 2 F_1^* F_2 + \cos \Theta
\left( \mid F_1 \mid^2 + \mid F_2 \mid^2 \right)
\right. \nonumber \\
& & \left. - \sin^2 \Theta \left( F_1^* F_3 + F_2^* F_4 \right) \right\}.
\nonumber
\end{eqnarray}

\section{Response Functions and Helicity Amplitudes}
\label{helapp}
\begin{eqnarray}
R_T^{00} & = & \frac{1}{2} \left(
\mid H_1 \mid^2
+\mid H_2 \mid^2
+\mid H_3 \mid^2
+\mid H_4 \mid^2
\right),
\nonumber
\\
R_T^{0y} & = & - \Im \left\{ H_1^* H_2 + H_3^* H_4 \right\},
\nonumber
\\
R_T^{y'0} & = & \Im \left\{ H_1^* H_3 + H_2^* H_4 \right\},
\nonumber
\\
R_T^{x' x} & = & \Re \left\{ H_1^* H_4 + H_2^* H_3 \right\},
\nonumber
\\
R_T^{x' z} & = & \Re \left\{ H_1^* H_3 - H_2^* H_4 \right\},
\nonumber
\\
R_T^{z' x} & = & \Re \left\{ H_1^* H_2 - H_3^* H_4 \right\},
\nonumber
\\
R_T^{z' z} & = & \frac{1}{2}
\left(
\mid H_1 \mid^2 - \mid H_2 \mid^2 - \mid H_3 \mid^2 + \mid H_4 \mid^2
\right),
\nonumber
\\
R_L^{00} & = & \mid H_5 \mid^2 + \mid H_6 \mid^2,
\nonumber
\\
R_L^{0y} & = & - 2 \Im \left\{ H_5^* H_6 \right\},
\nonumber
\\
R_L^{x' x} & = & - \mid H_5 \mid^2 + \mid H_6 \mid^2,
\nonumber
\\
R_L^{z' x} & = & 2 \Re \left\{ H_5^* H_6 \right\},
\nonumber
\\
{ }^c \! R_{TL}^{00} & = & \frac{1}{\sqrt{2}} \Re
\left\{ H_5^* H_1 - H_5^* H_4 + H_6^* H_2 + H_6^* H_3 \right\},
\nonumber
\\
{ }^s \! R_{TL}^{0x} & = & \frac{1}{\sqrt{2}} \Im
\left\{ - H_5^* H_2 + H_5^* H_3 - H_6^* H_1 - H_6^* H_4 \right\},
\nonumber
\\
{ }^c \! R_{TL}^{0y} & = & \frac{1}{\sqrt{2}} \Im
\left\{ - H_5^* H_2 - H_5^* H_3 + H_6^* H_1 - H_6^* H_4 \right\},
\nonumber
\\
{ }^s \! R_{TL}^{0z} & = & \frac{1}{\sqrt{2}} \Im
\left\{ - H_5^* H_1 - H_5^* H_4 + H_6^* H_2 - H_6^* H_3 \right\},
\nonumber
\\
{ }^s \! R_{TL}^{x'0} & = & \frac{1}{\sqrt{2}} \Im
\left\{ H_5^* H_2 - H_5^* H_3 - H_6^* H_1 - H_6^* H_4 \right\},
\nonumber
\\
{ }^s \! R_{TL}^{z'0} & = & \frac{1}{\sqrt{2}} \Im
\left\{ - H_5^* H_1 - H_5^* H_4 - H_6^* H_2 + H_6^* H_3 \right\},
\nonumber
\\
{ }^c \! R_{TL}^{x'x} & = & \frac{1}{\sqrt{2}} \Re
\left\{ - H_5^* H_1 + H_5^* H_4 + H_6^* H_2 + H_6^* H_3 \right\},
\nonumber
\\
{ }^c \! R_{TL}^{z'x} & = & \frac{1}{\sqrt{2}} \Re
\left\{ H_5^* H_2 + H_5^* H_3 + H_6^* H_1 - H_6^* H_4 \right\},
\nonumber
\\
{ }^c \! R_{TT}^{00} & = & \Re \left\{ - H_1^* H_4 + H_2^* H_3 \right\},
\nonumber
\\
{ }^s \! R_{TT}^{0x} & = & \Im \left\{ H_1^* H_3 - H_2^* H_4 \right\},
\nonumber
\\
{ }^s \! R_{TT}^{0z} & = & - \Im \left\{ H_1^* H_4 + H_2^* H_3 \right\},
\nonumber
\\
{ }^s \! R_{TT}^{x'0} & = & \Im \left\{ H_1^* H_2 - H_3^* H_4 \right\},
\nonumber
\\
{ }^s \!R_{TT}^{z'0} & = & \Im \left\{ - H_1^* H_4 + H_2^* H_3 \right\},
\nonumber
\\
{ }^s \! R_{TL'}^{00} & = & \frac{1}{\sqrt{2}} \Im
\left\{ - H_5^* H_1 + H_5^* H_4 - H_6^* H_2 - H_6^* H_3 \right\},
\nonumber
\\
{ }^c \! R_{TL'}^{0x} & = & \frac{1}{\sqrt{2}} \Re
\left\{ H_5^* H_2 - H_5^* H_3 + H_6^* H_1 + H_6^* H_4 \right\},
\nonumber
\\
{ }^s \! R_{TL'}^{0y} & = & \frac{1}{\sqrt{2}} \Re
\left\{ - H_5^* H_2 - H_5^* H_3 + H_6^* H_1 - H_6^* H_4 \right\},
\nonumber
\\
{ }^c \! R_{TL'}^{0z} & = & \frac{1}{\sqrt{2}} \Re
\left\{ H_5^* H_1 + H_5^* H_4 - H_6^* H_2 + H_6^* H_3 \right\},
\nonumber
\\
{ }^c \! R_{TL'}^{x'0} & = & \frac{1}{\sqrt{2}} \Re
\left\{ - H_5^* H_2 + H_5^* H_3 + H_6^* H_1 + H_6^* H_4 \right\},
\nonumber
\\
{ }^c \! R_{TL'}^{z'0} & = & \frac{1}{\sqrt{2}} \Re
\left\{ H_5^* H_1 + H_5^* H_4 + H_6^* H_2 - H_6^* H_3 \right\},
\nonumber
\\
{ }^s \! R_{TL'}^{x'x} & = & \frac{1}{\sqrt{2}} \Im
\left\{ H_5^* H_1 - H_5^* H_4 - H_6^* H_2 - H_6^* H_3 \right\},
\nonumber
\\
{ }^s \! R_{TL'}^{z'x} & = & \frac{1}{\sqrt{2}} \Im
\left\{ - H_5^* H_2 - H_5^* H_3 - H_6^* H_1 + H_6^* H_4 \right\},
\nonumber
\\
R_{TT'}^{0x} & = & \Re \left\{ H_1^* H_2 + H_3^* H_4 \right\},
\nonumber
\\
R_{TT'}^{0z} & = & \frac{1}{2} \left( \mid H_1 \mid^2 - \mid H_2 \mid^2
+ \mid H_3 \mid^2 - \mid H_4 \mid^2 \right),
\nonumber
\\
R_{TT'}^{x'0} & = & \Re \left\{ H_1^* H_3 + H_2^* H_4 \right\},
\nonumber
\\
R_{TT'}^{z'0} & = & \frac{1}{2}
\left( \mid H_1 \mid^2 + \mid H_2 \mid^2
- \mid H_3 \mid^2 - \mid H_4 \mid^2 \right).
\nonumber
\end{eqnarray}

\section[Multipole Expansion of Response Functions]
{Multipole Expansion of Response Functions for
Eta Electroproduction}
\label{multapp}
In the following we have kept all terms proportional to the dominant
multipoles $E_{0+}$ and $L_{0+}$ and interference terms with the smaller
multipoles considered in the text,
$M_{1-}$, $L_{1-}$, $E_{2-}$, $M_{2-}$ and $L_{2-}$.
\begin{eqnarray}
R_T^{00} & = & \mid E_{0+} \mid^2
- \Re \left\{ E_{0+}^* \left[ 2 \cos \Theta M_{1-}
\right. \right. \nonumber \\
& & \left. \left. -
\left( 3 \cos^2 \Theta - 1 \right)
\left( E_{2-} - 3 M_{2-} \right) \right] \right\}, \nonumber
\\
R_{T}^{0y} & = &
- 3 \sin \Theta \cos \Theta \Im \left\{ E_{0+}^* \left( E_{2-} + M_{2-}
\right) \right\}
, \nonumber
\\
R_{T}^{y'0} & = &
- \sin \Theta \Im \left\{ E_{0+}^* \left[ 2 M_{1-} - 3 \cos \Theta
\left( E_{2-} - 3 M_{2-} \right) \right] \right\}
, \nonumber
\\
R_{T}^{x'x} & = &
0
, \nonumber
\\
R_{T}^{x'z} & = &
\sin \Theta \left[ \mid E_{0+} \mid^2 - \Re \left\{ E_{0+} \left(
E_{2-} - 3 M_{2-} \right) \right\} \right]
, \nonumber
\\
R_{T}^{z'x} & = &
- 3 \sin \Theta \Re \left\{ E_{0+}^* \left( E_{2-} + M_{2-} \right) \right\}
, \nonumber
\\
R_{T}^{z'z} & = &
- \cos \Theta \mid E_{0+} \mid^2 + 2 \Re \left\{ E_{0+}^* \left[
M_{1-} \right. \right. \nonumber \\
& & \left. \left.
- \cos \Theta \left( E_{2-} - 3 M_{2-} \right) \right] \right\}
, \nonumber
\\
R_{L}^{00} & = &
\mid L_{0+} \mid^2
+ 2 \Re \left\{ L_{0+}^* \left(
\cos \Theta L_{1-} \right. \right. \nonumber \\
& & \left. \left. - 2 \left( 1 - 3 \cos^2 \Theta \right)
L_{2-} \right) \right\}
, \nonumber
\\
R_{L}^{0y} & = &
- 2 \sin \Theta \Im \left\{ L_{0+}^* \left( L_{1-} + 6 \cos \Theta L_{2-}
\right) \right\}
, \nonumber
\\
R_{L}^{x'x} & = &
- \cos \Theta \mid L_{0+} \mid^2
- 2 \Re \left\{ L_{0+}^* \left( L_{1-} + 4 \cos \Theta L_{2-} \right) \right\}
, \nonumber
\\
R_{L}^{z'x} & = &
- \sin \Theta ( \mid L_{0+} \mid^2 + 4 \Re \left\{ L_{0+}^* L_{2-} \right\})
, \nonumber
\\
{ }^c \! R_{TL}^{00} & = &
- \sin \Theta \Re \left\{ E_{0+}^*
\left( L_{1-} + 6 \cos \Theta L_{2-} \right) \right. \nonumber \\
& & \left.
+ L_{0+}^* \left( M_{1-} + 3 \cos \Theta \left( M_{2-} - E_{2-} \right)
\right) \right\}
, \nonumber
\\
{ }^s \! R_{TL}^{0x} & = &
\Im \left\{ L_{0+}^* E_{0+}
+ E_{0+}^* \left( - \cos \Theta L_{1-} \right. \right. \nonumber \\
& & \left. \left. + 2 \left( 1 - 3  \cos^2 \Theta L_{2-}
\right)
\right)
+ L_{0+}^* \left( - \cos \Theta
M_{1-} \right. \right. \nonumber \\
& & \left. \left.
+ E_{2-} + 3 \left( 1 - 2 \cos^2 \Theta \right) M_{2-} \right)
\right\}
, \nonumber
\\
{ }^c \! R_{TL}^{0y} & = &
\Im \left\{ L_{0+}^* E_{0+}
+ E_{0+}^* \left( - \cos \Theta L_{1-} \right. \right. \nonumber \\
& & \left. \left. + 2 \left( 1 - 3 \cos^2 \Theta
\right) L_{2-} \right)
+ L_{0+}^* \left( - \cos \Theta M_{1-} \right. \right. \nonumber \\
& & \left. \left. + \left( 3 \cos^2 \Theta - 2 \right)
E_{2-} - 3 \cos^2 \Theta M_{2-} \right) \right\}
, \nonumber
\\
{ }^s \! R_{TL}^{0z} & = &
\sin \Theta \Im \left\{ E_{0+}^* \left( L_{1-} + 6 \cos \Theta L_{2-} \right)
\right. \nonumber \\
& & \left. - L_{0+}^* \left( M_{1-} + 6 \cos \Theta M_{2-} \right) \right\}
, \nonumber
\\
{ }^s \! R_{TL}^{x'0} & = &
\Im \left\{ - \cos \Theta L_{0+}^* E_{0+}
+ E_{0+}^* \left( L_{1-} + 4 \cos \Theta L_{2-} \right) \right. \nonumber \\
& & \left. + L_{0+}^*
\left( M_{1-} + \cos \Theta \left( 3 M_{2-} - E_{2-} \right)
\right) \right\}
, \nonumber
\\
{ }^s \! R_{TL}^{z'0} & = &
- \sin \Theta \Im \left\{ L_{0+}^* E_{0+} + 2 E_{0+}^* L_{2-}
\right. \nonumber \\
& & \left. + L_{0+}^* \left( E_{2-} + 3 M_{2-} \right) \right\}
, \nonumber
\\
{ }^c \! R_{TL}^{x'x} & = &
\sin \Theta \Re \left\{ L_{0+} E_{0+} - 2 E_{0+} L_{2-} - 2 L_{0+} E_{2-}
\right\}
, \nonumber
\\
{ }^c \! R_{TL}^{z'x} & = &
\Re \left\{ - \cos \Theta L_{0+}^* E_{0+}
- E_{0+}^* \left( L_{1-} + 4 \cos \Theta L_{2-} \right) \right. \nonumber \\
& & \left. + L_{0+}^* \left( M_{1-} - \cos \Theta \left( E_{2-} -
3 M_{2-} \right)
\right) \right\}
, \nonumber
\\
{ }^c \! R_{TT}^{00} & = & - 3
\sin^2 \Theta \Re \left\{ E_{0+}^* \left( E_{2-} + M_{2-}
\right) \right\}
, \nonumber
\\
{ }^s \! R_{TT}^{0x} & = &
\sin \Theta \Im \left\{ E_{0+}^* \left[ 2 M_{1-} - 3 \cos \Theta
\left( E_{2-} - 3 M_{2-} \right) \right] \right\}
, \nonumber
\\
{ }^s \! R_{TT}^{0z} & = &
3 \sin^2 \Theta \Im \left\{ E_{0+}^* \left( E_{2-} + M_{2-} \right)
\right\}
, \nonumber
\\
{ }^s \! R_{TT}^{x'0} & = &
3 \sin \Theta \Im \left\{ E_{0+}^* \left( E_{2-} + M_{2-} \right) \right\}
, \nonumber
\\
{ }^s \! R_{TT}^{z'0} & = &
0
, \nonumber
\\
{ }^s \! R_{TL'}^{00} & = &
\sin \Theta \Im \left\{
E_{0+}^* \left( - L_{1-} - 6 \cos \Theta L_{2-} \right) \right. \nonumber \\
& & \left.
+ L_{0+}^* \left( M_{1-} + 3 \cos \Theta \left( M_{2-} - E_{2-} \right)
\right) \right\}
, \nonumber
\\
{ }^c \! R_{TL'}^{0x} & = &
\Re \left\{ - L_{0+}^* E_{0+}
+ E_{0+}^* \left( - \cos \Theta L_{1-} \right. \right. \nonumber \\
& & \left. \left. + 2 \left( 1 - 3 \cos^2 \Theta \right)
L_{2-} \right)
\right. \nonumber \\
& & \left.
+ L_{0+}^* \left( \cos \Theta M_{1-} - 3 \left( 1 - 2 \cos^2 \Theta \right)
M_{2-} - E_{2-} \right) \right\}
, \nonumber
\\
{ }^s \! R_{TL'}^{0y} & = &
\Re \left\{ L_{0+}^* E_{0+}
+ E_{0+}^* \left( \cos \Theta L_{1-} \right. \right. \nonumber \\
& & \left. \left. - 2 \left( 1 - 3 \cos^2 \Theta
\right) L_{2-} \right) \right. \nonumber \\
& & \left.
+ L_{0+}^* \left( - \cos \Theta M_{1-}
+ \left( 1 - 3 \sin^2 \Theta \right) E_{2-}
\right. \right. \nonumber \\
& & \left. \left. - 3 \cos^2 \Theta M_{2-} \right) \right\}
, \nonumber
\\
{ }^c \!R_{TL'}^{0z} & = &
\sin \Theta \Re \left\{ E_{0+}^* \left( L_{1-} + 6 \cos \Theta L_{2-} \right)
\right. \nonumber \\
& & \left. + L_{0+}^* \left( M_{1-} + 6 \cos \Theta M_{2-} \right) \right\}
, \nonumber
\\
{ }^c \! R_{TL'}^{x'0} & = &
\Re \left\{ \cos \Theta L_{0+}^* E_{0+}
+ E_{0+}^* \left( L_{1-} + 4 \cos \Theta L_{2-} \right) \right. \nonumber \\
& & \left. + L_{0+}^* \left( - M_{1-} + \cos \Theta \left( E_{2-} - 3 M_{2-}
\right) \right) \right\}
, \nonumber
\\
{ }^c \! R_{TL'}^{z'0} & = &
\sin \Theta \Re
\left\{ L_{0+}^* E_{0+}
- 2 E_{0+}^* L_{2-} \right. \nonumber \\
& & \left.
+ L_{0+} \left( E_{2-} + 3 M_{2-} \right) \right\}
, \nonumber
\\
{ }^s \! R_{TL'}^{x'x} & = &
\sin \Theta \Im \left\{ - L_{0+}^* E_{0+}
- 2 E_{0+}^* L_{2-} + 2 L_{0+}^* E_{2-} \right\}
, \nonumber
\\
{ }^s \! R_{TL'}^{z'x} & = &
\Im \left\{ \cos \Theta L_{0+}^* E_{0+}
+ E_{0+}^* \left( - L_{1-} - 4 \cos \Theta L_{2-} \right) \right. \nonumber \\
& & \left. + L_{0+}^* \left( - M_{1-} +
\cos \Theta \left( E_{2-} - 3 M_{2-} \right)
\right) \right\}
, \nonumber
\\
R_{TT'}^{0x} & = &
- 3 \sin \Theta \cos \Theta \Re \left\{ E_{0+}^* \left( E_{2-} + M_{2-}
\right)
\right\}
, \nonumber
\\
R_{TT'}^{0z} & = &
- \mid E_{0+} \mid^2 + \Re
\left\{
E_{0+}^* \left[ 2 \cos \Theta M_{1-}
\right. \right. \nonumber \\
& & \left. \left.
- \left( 3 \cos^2 \Theta - 1 \right) \left( E_{2-} - 3 M_{2-}
\right) \right] \right\}
, \nonumber
\\
R_{TT'}^{x'0} & = &
- \sin \Theta \left[ \mid E_{0+} \mid^2
- \Re \left\{ E_{0+}^* \left( E_{2-} - 3 M_{2-} \right) \right\} \right]
, \nonumber
\\
R_{TT'}^{z'0} & = &
\cos \Theta \mid E_{0+} \mid^2 - 2 \Re \left\{ E_{0+}^*
\left[ M_{1-} \right. \right. \nonumber \\
& & \left. \left. - \cos \Theta \left( E_{2-} - 3 M_{2-}
\right) \right] \right\}
. \nonumber
\end{eqnarray}

\newpage
\clearpage

\begin{figure}
\label{kinpic}
\end{figure}
\small
{\bf{Figure 1:}}
\small
Kinematics of an electroproduction experiment.


\begin{figure}
\label{polkoordinaten}
\end{figure}
{\bf{Figure 2:}}
\small
Frames for polarization vectors.


\begin{figure}[ht]
\label{cops1a1pn}
\end{figure}
{\bf{Figure 3:}}
\small Electromagnetic helicity couplings
of the resonance \protect{$S_{11}(1535)$} as function of momentum transfer.
The solid line is the calculation with
model M1 \protect{\cite{CKO69}}, the dotted line is the result of model
M2 \protect{\cite{War90}}, the dashed
line the calculation with model M3 \protect{\cite{KW90}}
and the dash-dotted line with model M4 \protect{\cite{CK94}}.


\begin{figure*}[ht]
\label{copd1a1pn}
\end{figure*}
{\bf{Figure 4:}}
\small Electromagnetic helicity couplings
of the resonance $D_{13}(1520)$ as function of momentum transfer.
For notation see Fig.
\protect{\ref{cops1a1pn}}.


\begin{figure*}[ht]
\label{copp1a1pn}
\end{figure*}
{\bf{Figure 5:}}
\small Electromagnetic helicity couplings
of the resonance $P_{11}(1440)$ as function of momentum transfer.
For notation see Fig.
\protect{\ref{cops1a1pn}}.


\begin{figure*}[ht]
\label{photowqinclp}
\end{figure*}
{\bf{Figure 6:}}
\small
Total cross section for photoproduction. The left figure shows the
inclusive cross section for eta photoproduction off the proton.
The TAPS data
\protect{\cite{Kr94}} are denoted by $\Box$, the Bonn data ($\circ$)
are from an
electroproduction experiment \protect{\cite{Wi93}}
at very low momentum transfer,
$Q^2 = 0.056 \, \, {\mathrm GeV}^2$.
In the energy regime
beyond the resonance we have included some old data ($\Diamond$ from
\protect{\cite{CBC68}},
$\ast$ from \protect{\cite{ABB68}}).
The solid line is the full calculation with resonances and background, the
dashed line is the calculation without background, the dotted line is the
calculation without $D_{13}(1520)$ and $P_{11}(1440)$ resonance and the
dash-dotted line is only the contribution of the non-resonant
background.
The right figure shows the same calculation for a neutron target.


\begin{figure}[h]
\label{ds752}
\end{figure}
{\bf{Figure 7:}}
\small
Angular distribution for eta production off the proton at
$\nu = 783 \, \, {\mathrm MeV}$, compared to the data of
the TAPS collaboration \protect{\cite{Kr94}}. The solid line is the
standard calculation, the long dashed line without Roper and the short dashed
line without $D_{13}$ resonance. The dotted line was calculated with
$\Gamma_{\eta} / \Gamma = 0.003$ and the dash-dotted line with
$\Gamma_{\eta} / \Gamma = 0.010$ for the $D_{13}$.


\begin{figure*}[ht]
\label{photowqincln}
\end{figure*}
{\bf{Figure 8:}}
\small
Differential cross section for eta production off
the proton (left) and the neutron (right)
as a function of excitation energy
$\nu$ in the lab frame and scattering angle $\Theta$ in the c.m. system.


\begin{figure*}[th]
\label{obs1p}
\label{obs2p}
\end{figure*}
{\bf{Figure 9:}}
\small
Excitation functions off the proton for
the unpolarized cross section $\sigma_0$,
the single polarization observables and the double polarization observables
for polarized beam and target at a scattering angle $\Theta = 90^o$ in the
c.m. system.
The solid line is our standard calculation, the
dotted line is without Roper resonance, the long dashed line
without $D_{13}(1520)$, the short dashed line without $D_{15}(1675)$
and the dash-dotted line
without non-resonant background. The data for $\sigma_0$ are from
\protect{\cite{Kr94}} (filled circles), \protect{\cite{Br94}} (boxes,
$\Theta = 95.7^o$), \protect{\cite{He}} (circles). The
recoil polarization data are from \protect{\cite{Bl}}.


\begin{figure*}[th]
\label{sigmal} \label{sl11}
\end{figure*}
{\bf{Figure 10:}}
\small
Transverse/longitudinal separation of the inclusive cross section.
The
left figure shows the longitudinal part $\sigma_L$
of the inclusive electroproduction cross
section on the proton. The data point is from
\protect{\cite{Ni78}}. The dotted curve is the result of a calculation
in the non-relativistic constituent quark model (M1)
\protect{\cite{CKO69,IK78}}. In the
calculation for the short dashed curve, the resonance part of the
\protect{$L_{0+}$} multipole
was calculated in the relativized constituent quark model (M2)
\protect{\cite{War90}}, the dash-dotted curve is the calculation in the
light cone model \protect{\cite{KW90}} and the long dashed line is the
contribution of the non-resonant background. The standard calculation
(solid line) has been performed in model M1 with the \protect{$L_{0+}$}
multipole normalized to
the data point. The right figure shows the longitudinal cross section,
$\varepsilon_L \sigma_L$ (dotted curve),
and the transverse cross section $\sigma_T$ (dashed curve)
in the standard calculation.
The data point at
$1533 \, \, {\mathrm MeV}$ is from
\protect{\cite{Ni78} }.


\begin{figure}[h]
\label{resp1p}
\end{figure}
{\bf{Figure 11:}}
\small
Unpolarized differential cross sections $d \sigma_{T}^{00} / d \Omega$
and $d \sigma_{L}^{00} / d \Omega$ at
$W = 1533 \, \, {\mathrm MeV}$. The left figures are calculated at
$Q^2 = 0.120 \, \, {\mathrm GeV}^2$, the right figures at
$Q^2 = 0.393 \, \, {\mathrm GeV}^2$.
The solid curve is our standard calculation (model M1),
for the short dashed line the resonant parts
of the $E_{0+}$ and $L_{0+}$ multipoles
were calculated in model M3, the dotted line is a calculation
without $P_{11}(1440)$, the long dashed line without
$D_{13}(1520)$, and the dash-dotted line without
non-resonant background.


\begin{figure}[h]
\label{resp2p}
\end{figure}
{\bf{Figure 12:}}
\small
Unpolarized differential cross sections $d \sigma_{TT}^{00} / d \Omega$
and $d \sigma_{TL}^{00} / d \Omega$ at
$W = 1533 \, \, {\mathrm MeV}$. The left figures were calculated at
$Q^2 = 0.120 \, \, {\mathrm GeV}^2$, the right figures at
$Q^2 = 0.393 \, \,
{\mathrm GeV}^2$. For notation see Fig. \protect{\ref{resp1p}}.


\begin{figure}[h]
\label{resp3p}
\end{figure}
{\bf{Figure 13:}}
\small
Differential cross sections $d \sigma_{TT}^{0x} / d \Omega$
and $d \sigma_{TT}^{0y} / d \Omega$ at
$W = 1533 \, \, {\mathrm MeV}$. The left figures were calculated at
$Q^2 = 0.120 \, \, {\mathrm GeV}^2$, the right figures at
$Q^2 = 0.393 \, \,
{\mathrm GeV}^2$. For notation see Fig. \protect{\ref{resp1p}}.

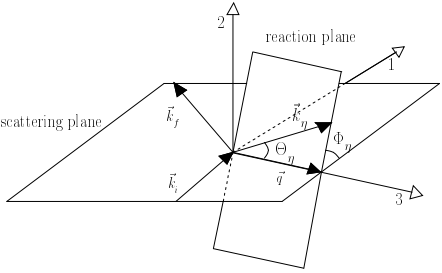
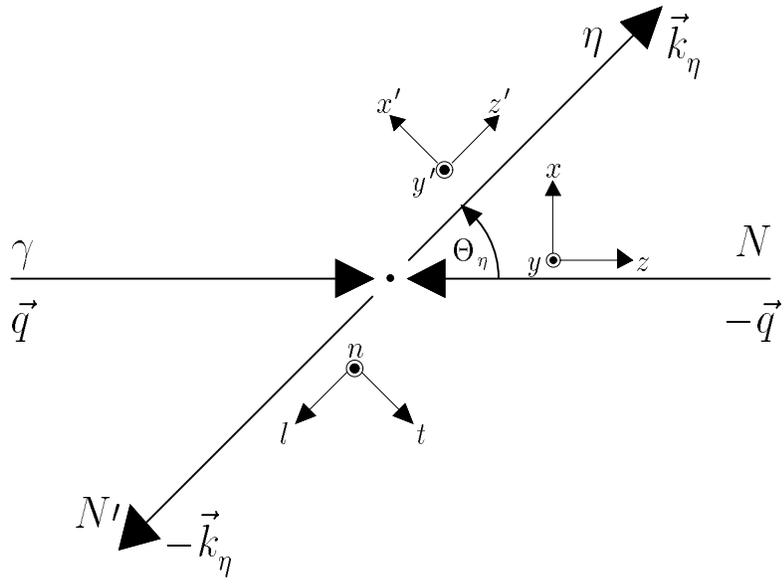
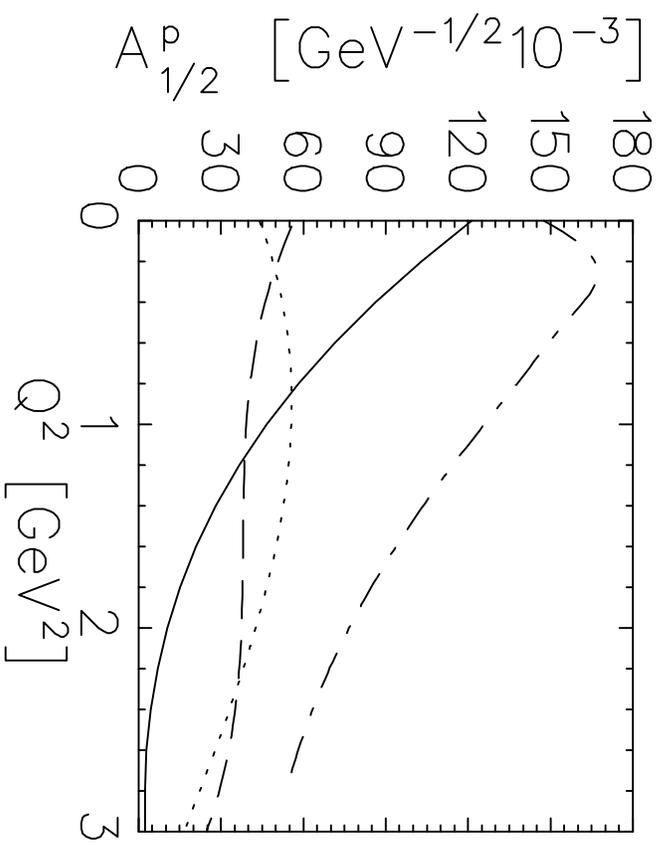
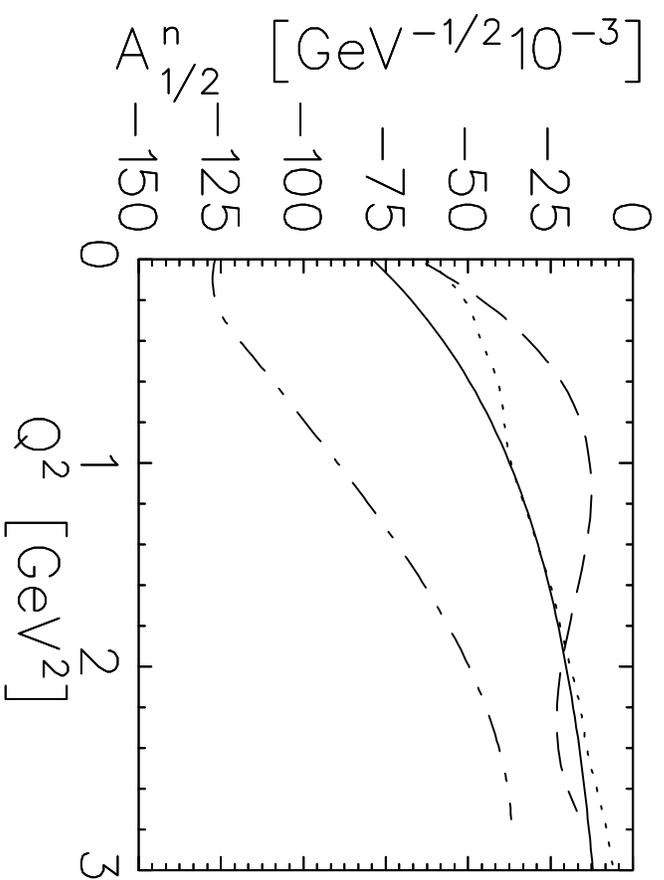
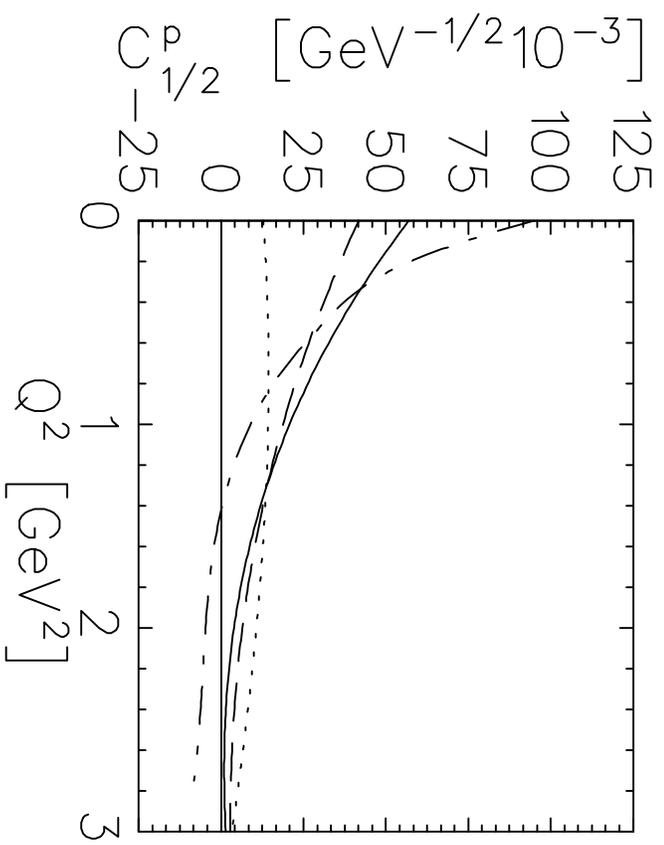
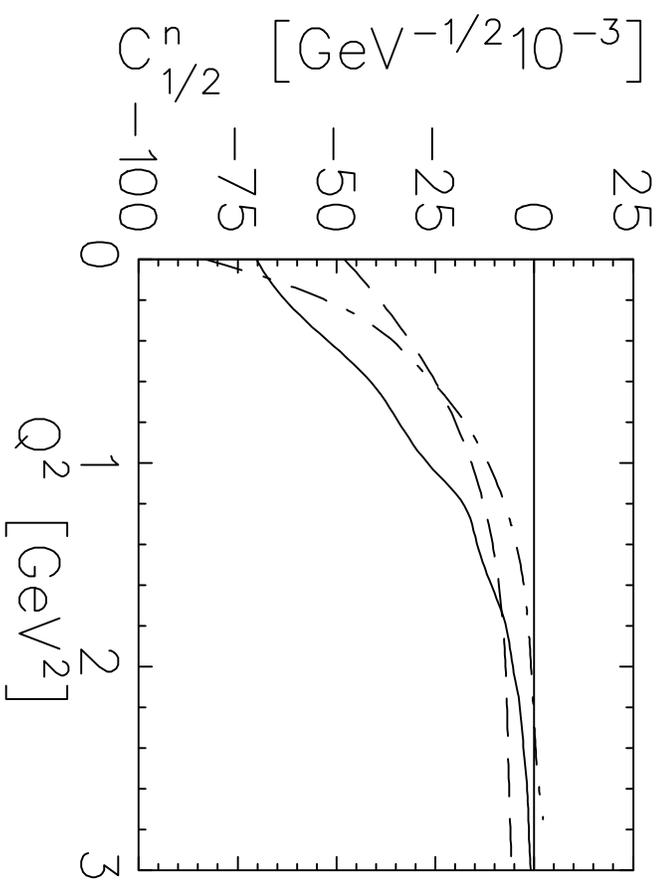
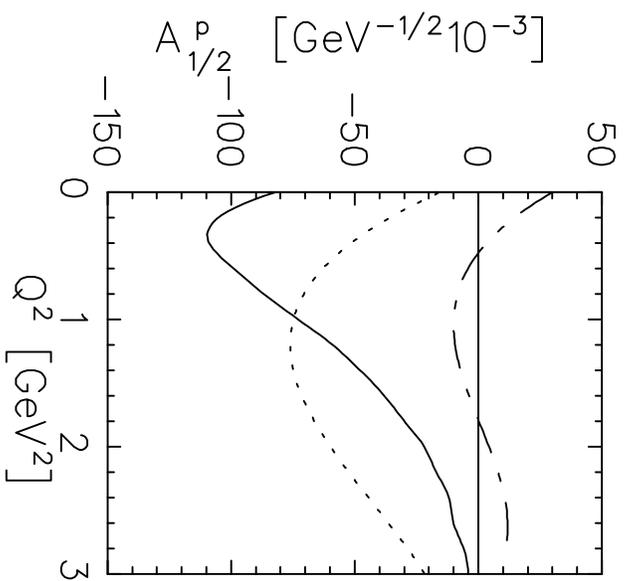
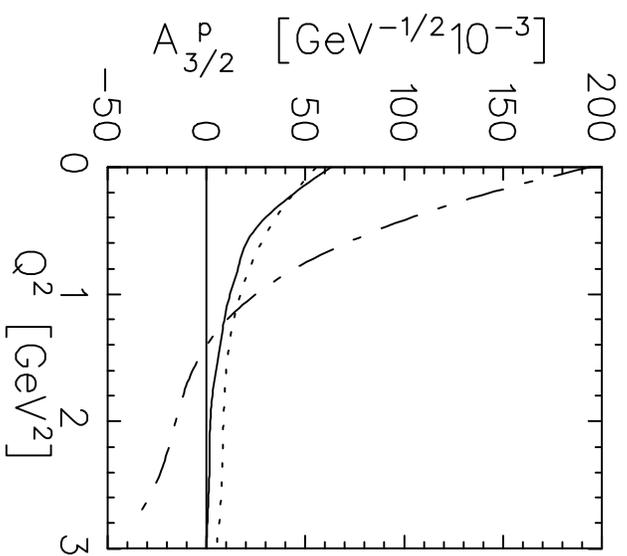
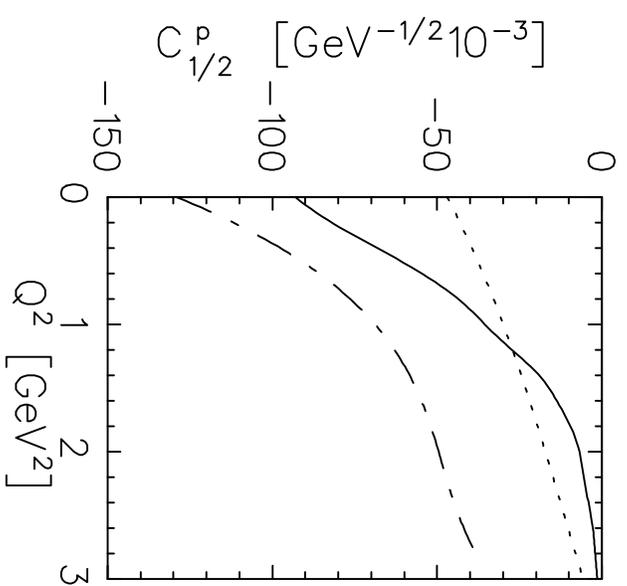
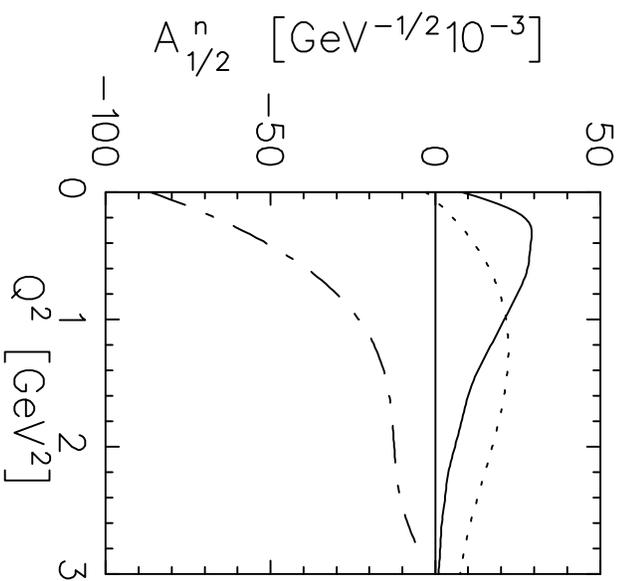
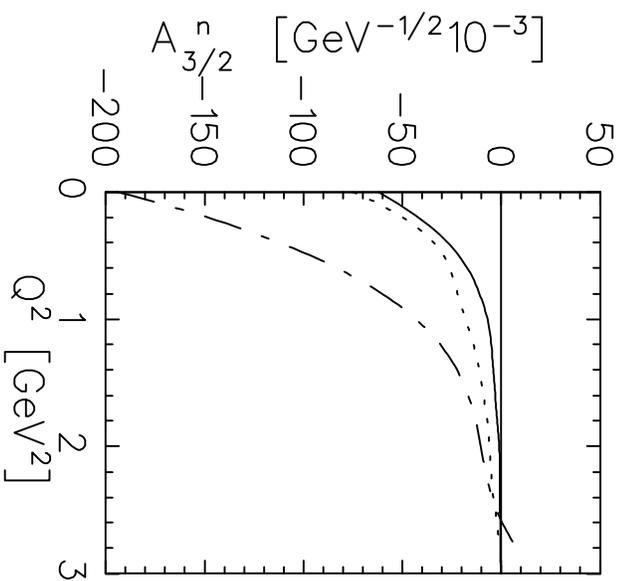
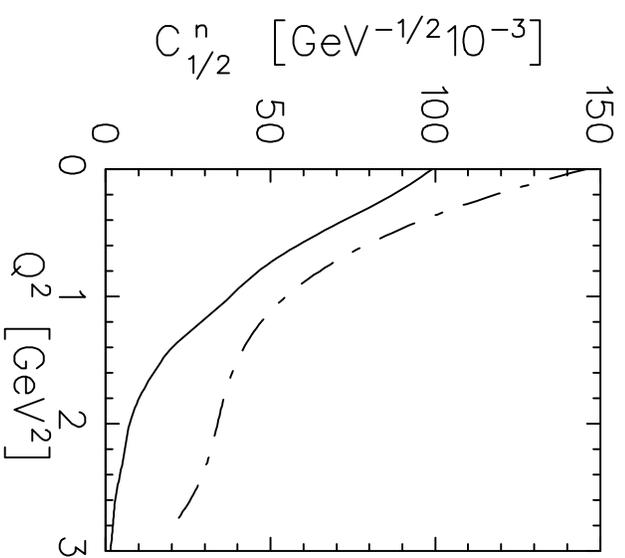
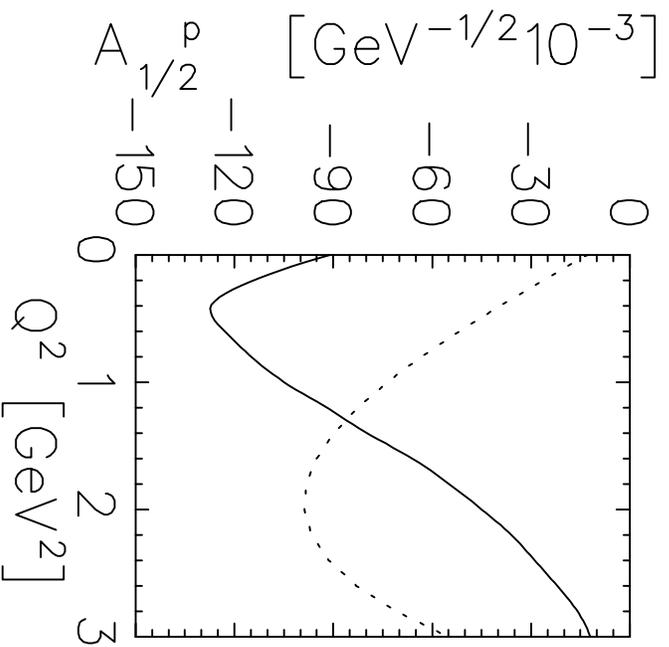
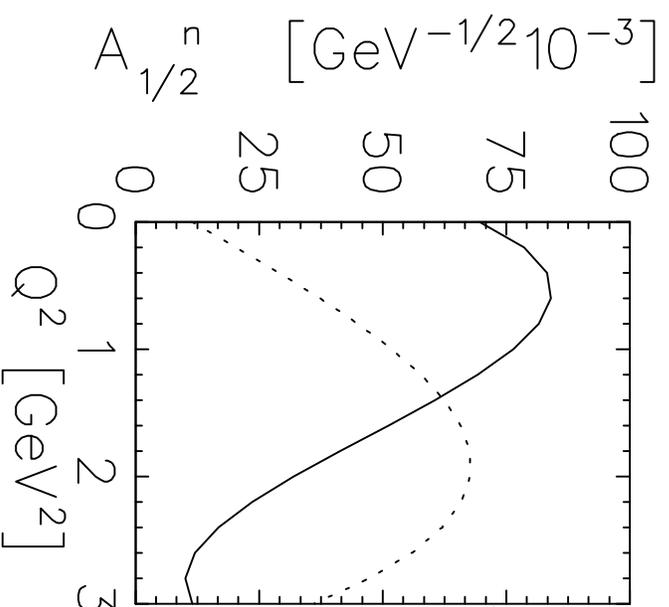
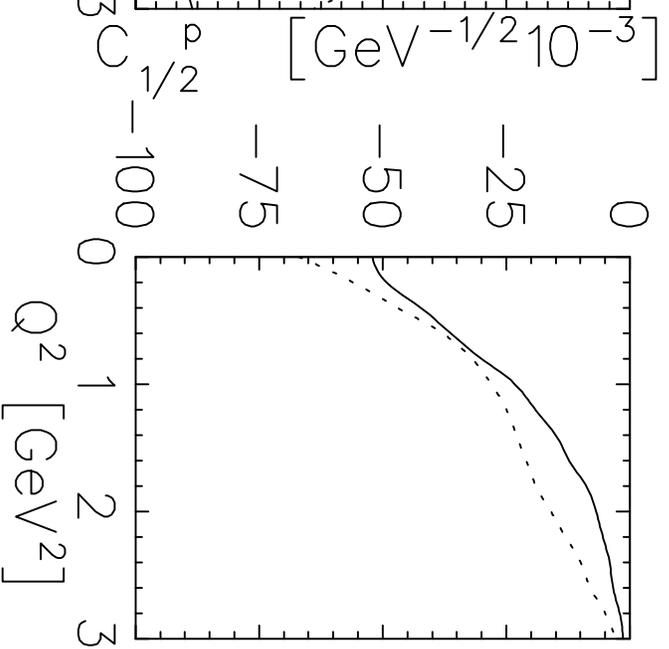
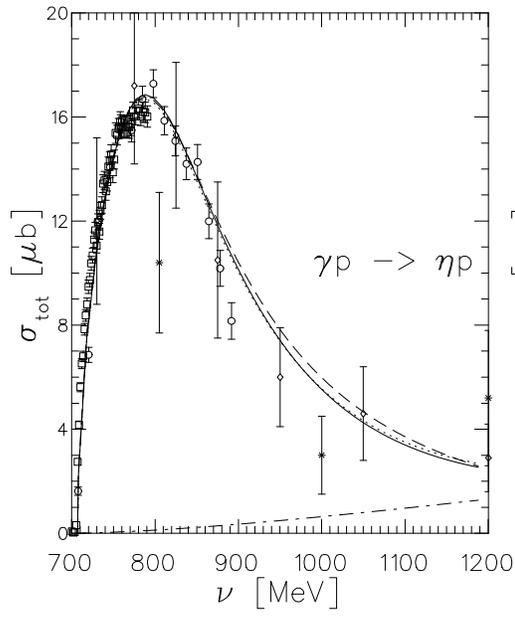
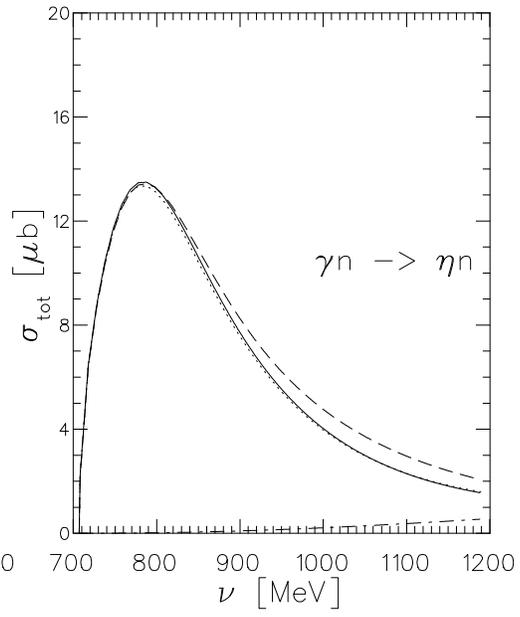
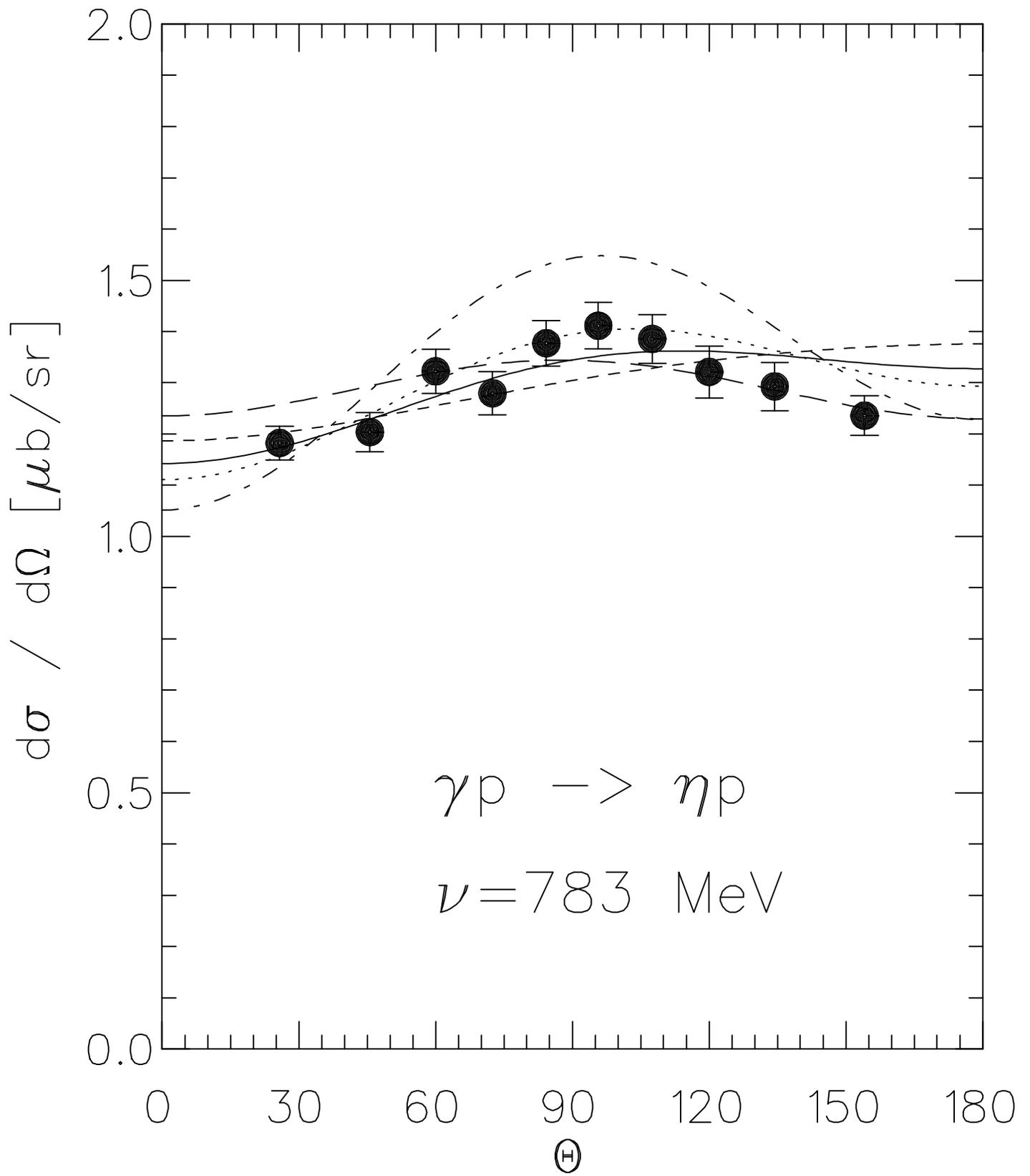
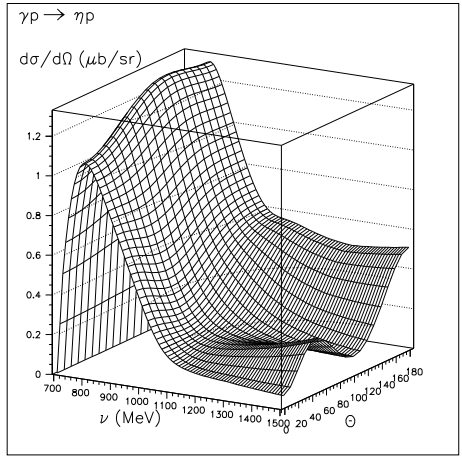
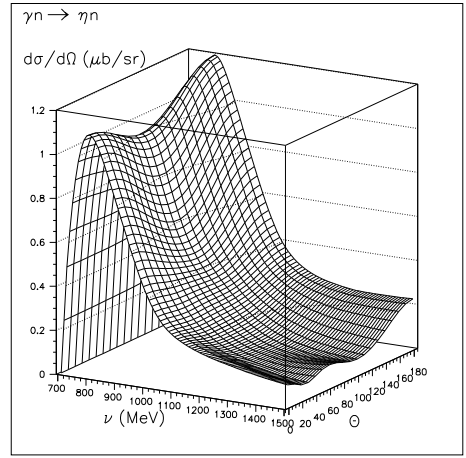
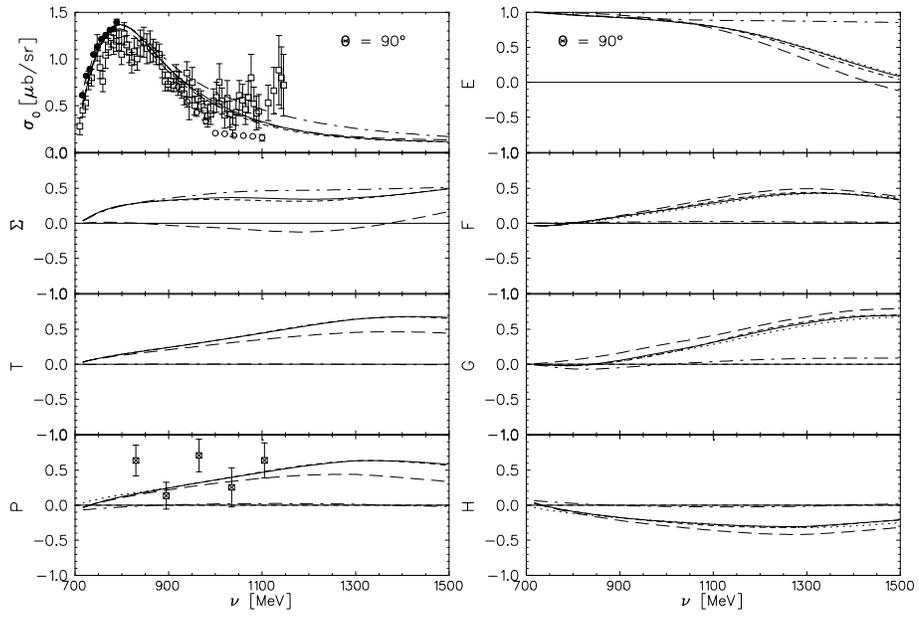
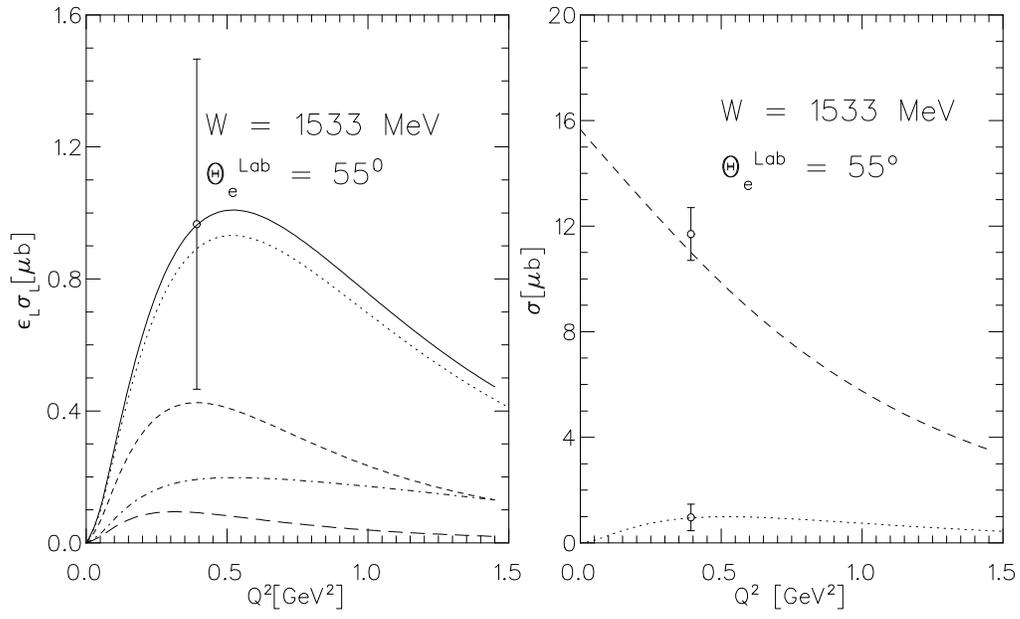
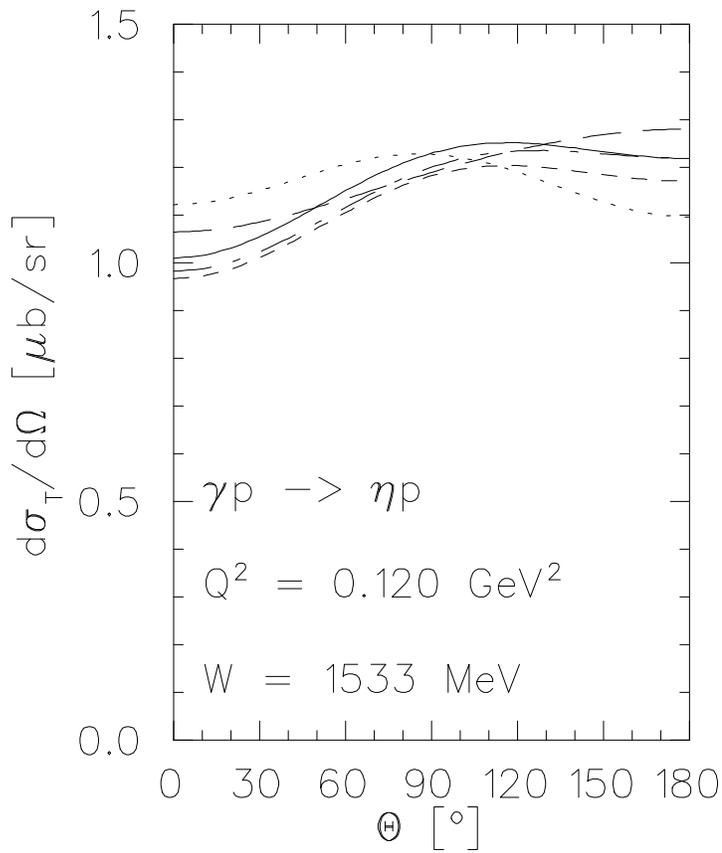
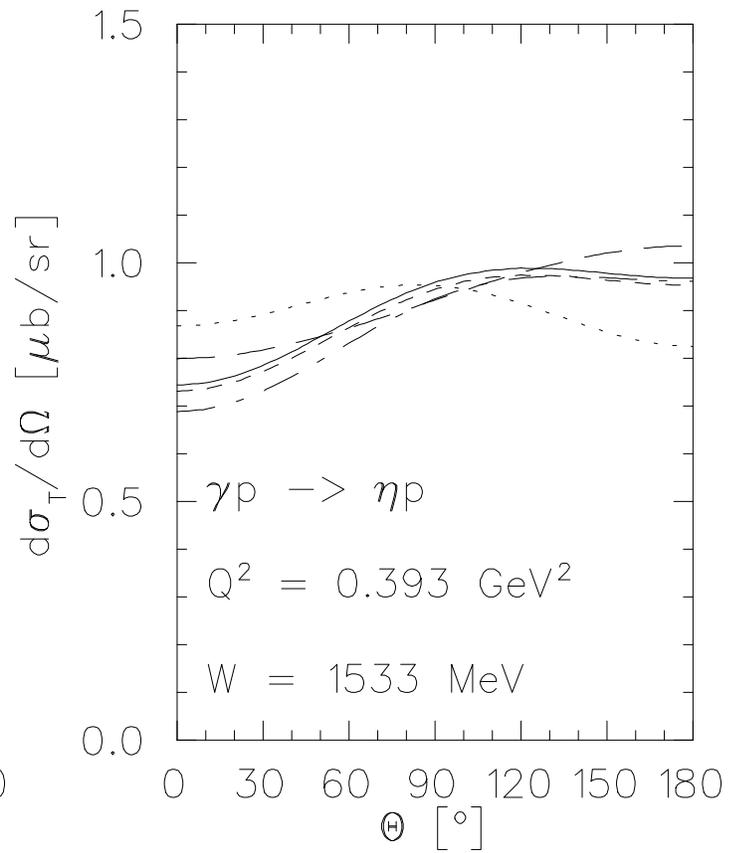
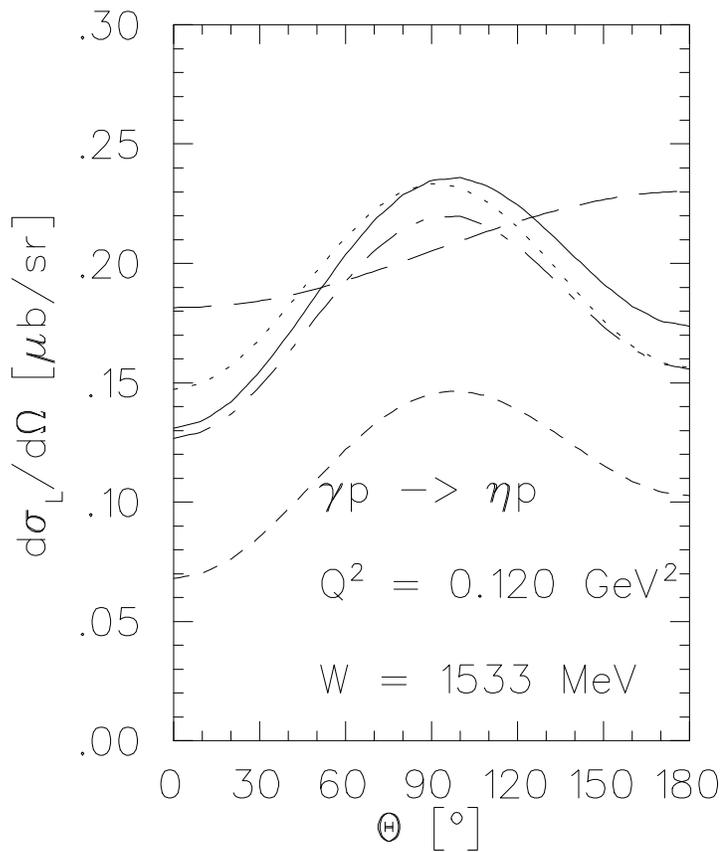
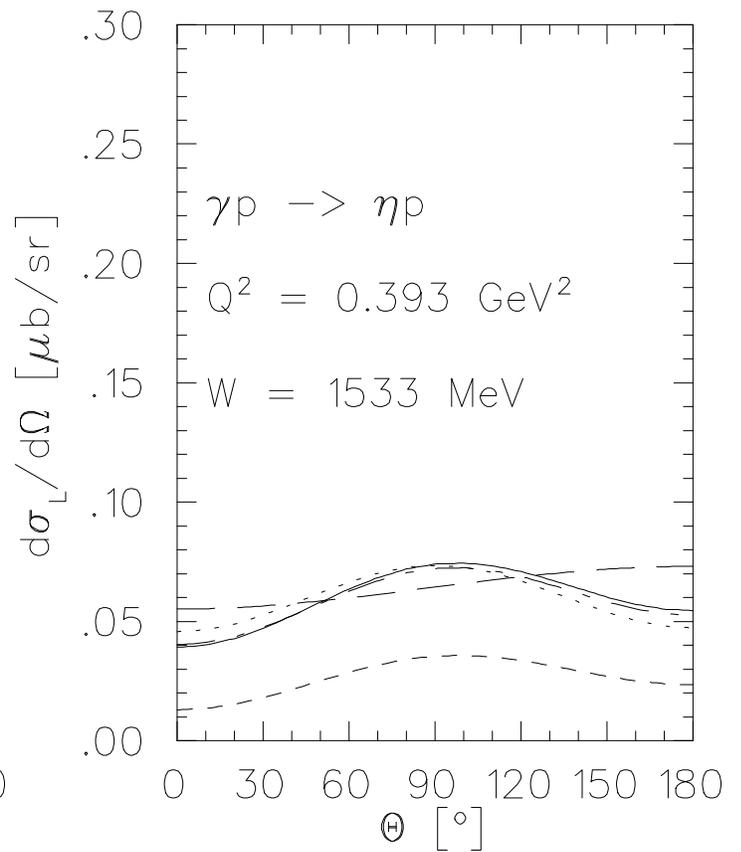
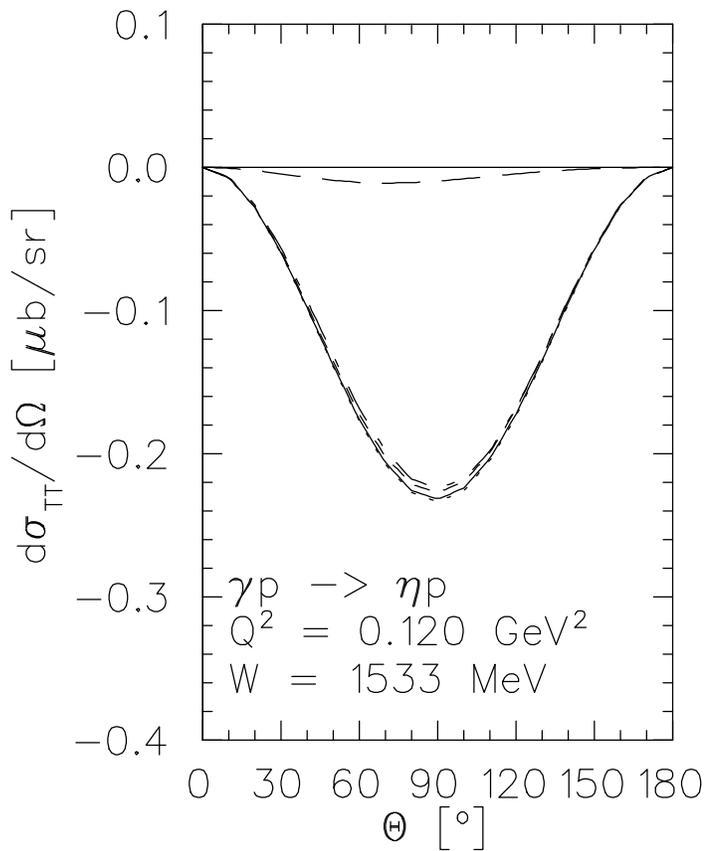
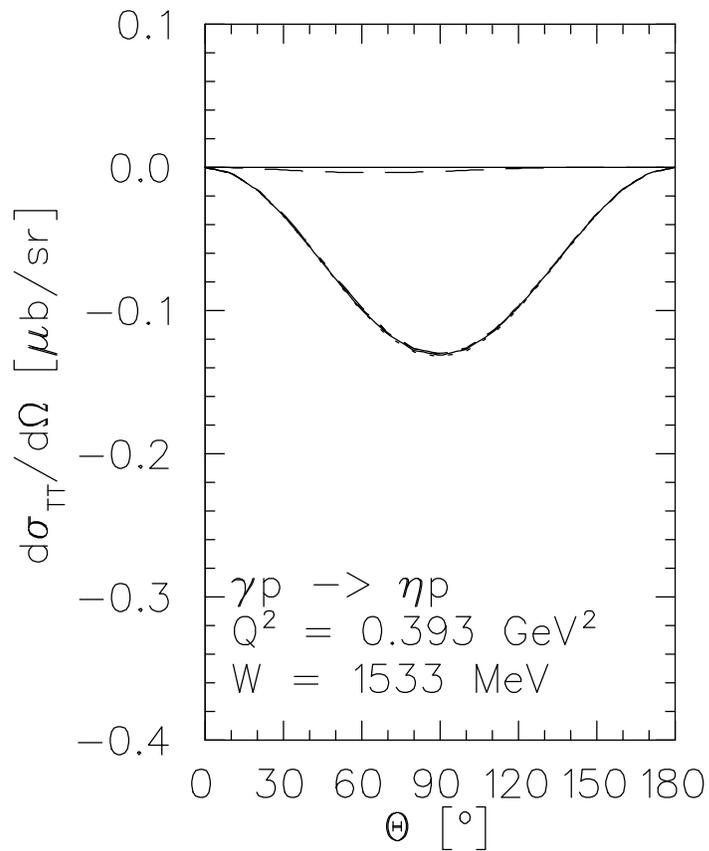
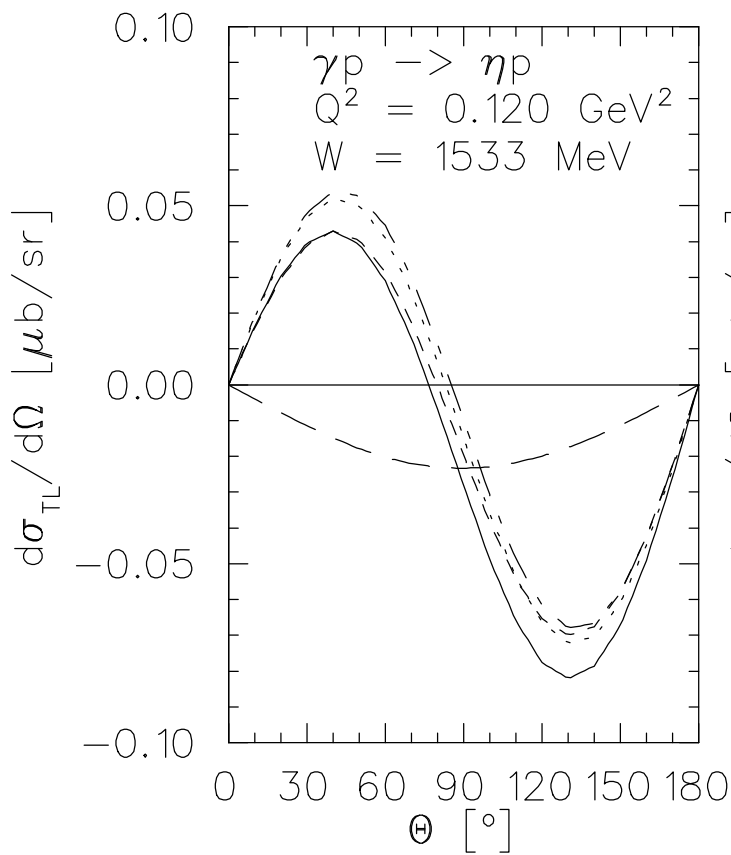
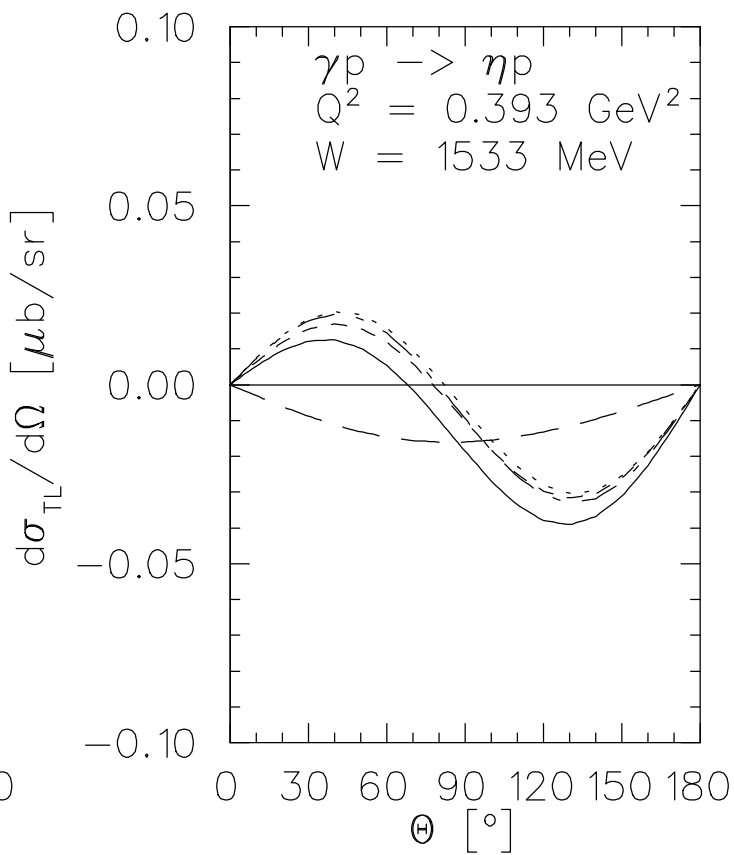
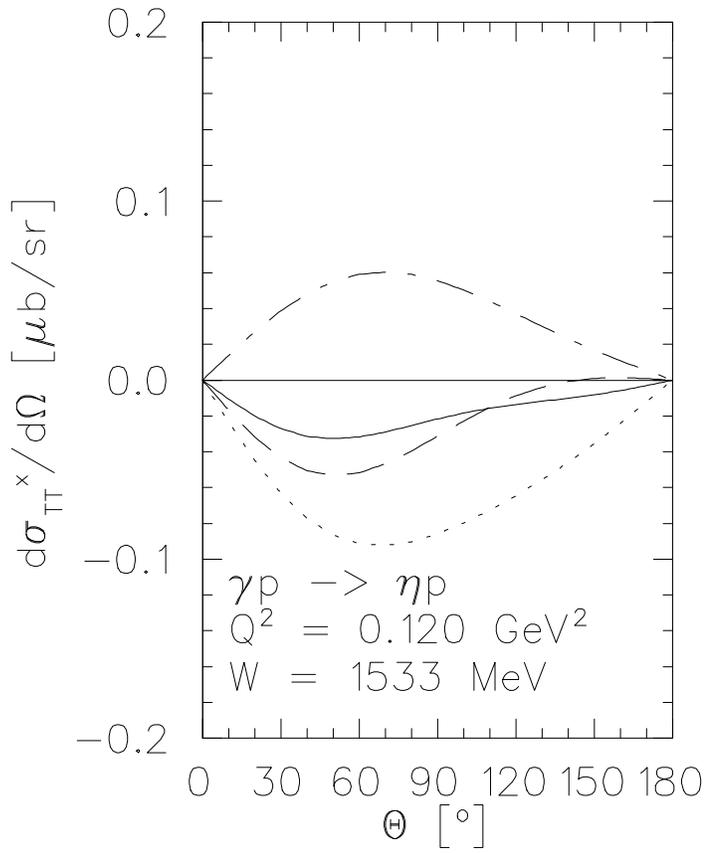
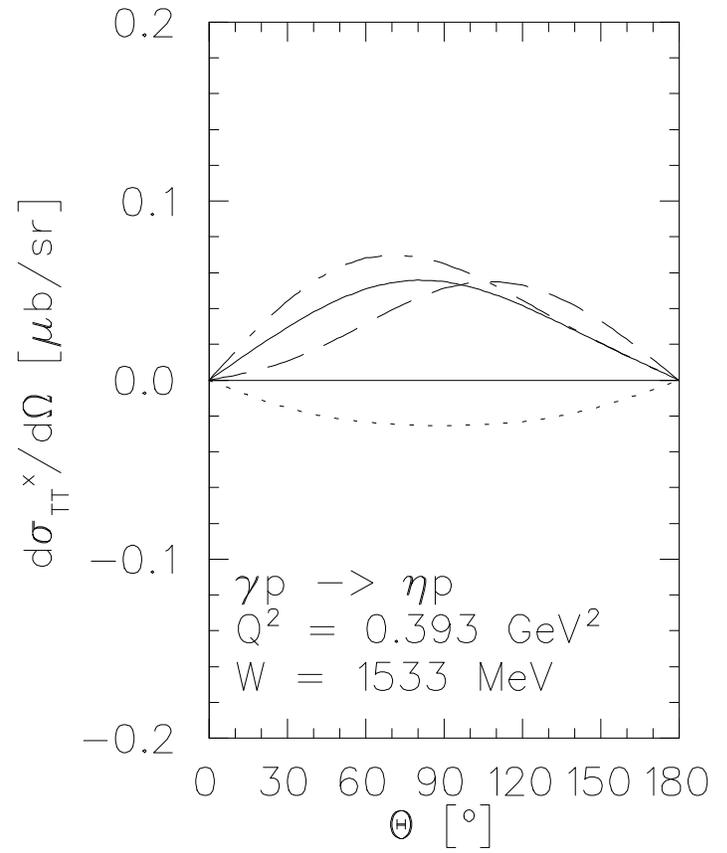
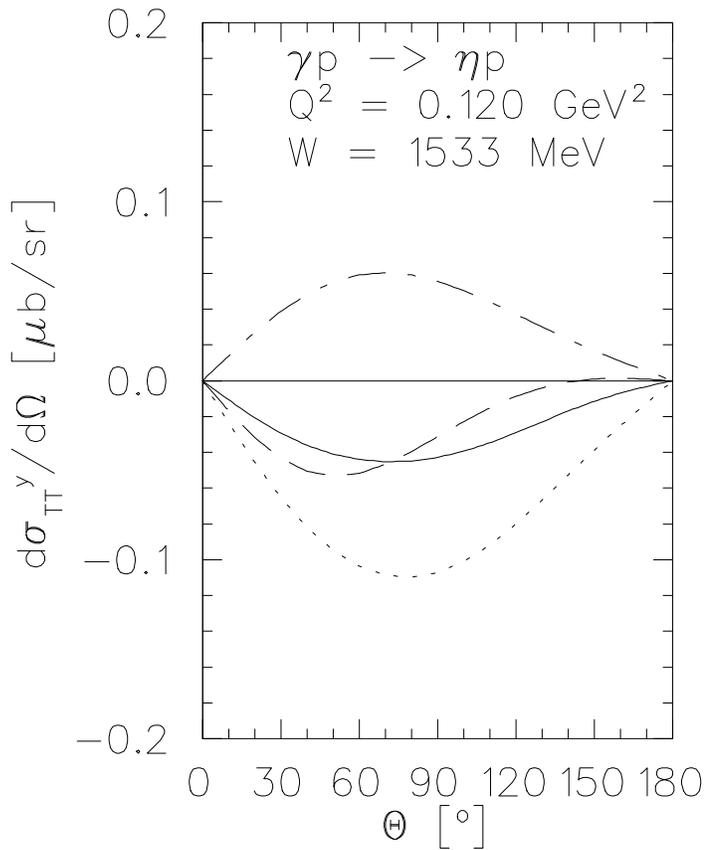
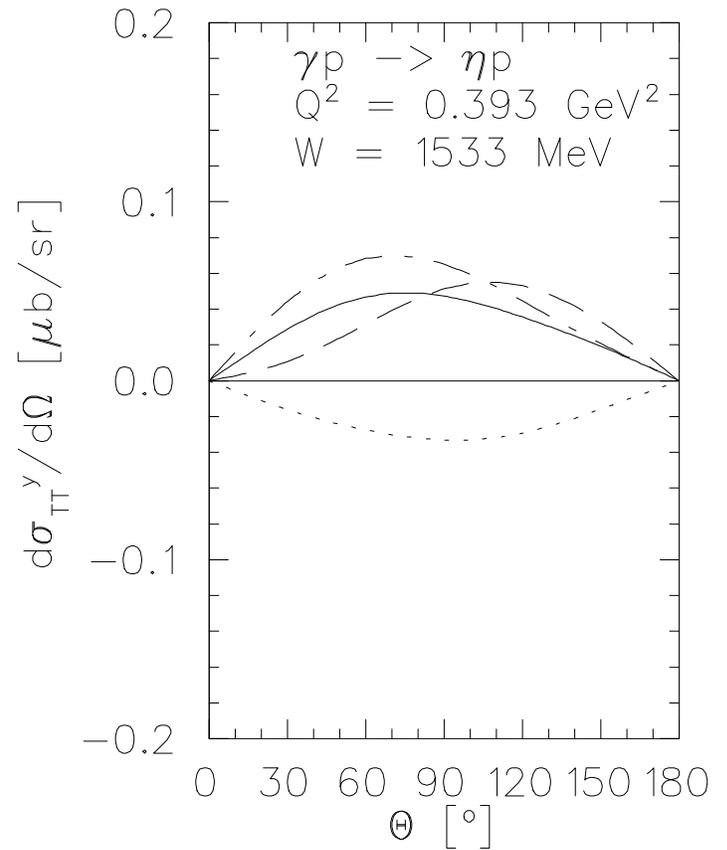
\end{document}